\documentclass[preprint,showpacs,preprintnumbers,amsmath,amssymb]{revtex4}

\usepackage{graphicx}
\usepackage{dcolumn}
\usepackage{bm}

\begin{document}


\title{Linear coupling and over-reflection phenomena of magnetohydrodynamic waves in smooth shear flows}

\author{G.~Gogoberidze}
\email{gogober@geo.net.ge}

\author{G.~D.~Chagelishvili}%
\affiliation{Center for Plasma Astrophysics, Abastumani
Astrophysical Observatory, Ave. A. Kazbegi 2, Tbilisi 0160,
Georgia}

\author{R.~Z.~Sagdeev}%
\affiliation{Department of Physics, University of Maryland,
College Park, Maryland 20740}

\author{D.~G.~Lominadze}%
\affiliation{Center for Plasma Astrophysics, Abastumani
Astrophysical Observatory, Ave. A.Kazbegi 2, Tbilisi 0160,
Georgia}

\date{\today}

\begin{abstract}

Special features of magnetohydrodynamic waves linear dynamics in
smooth  shear flows are studied. Quantitative asymptotic and
numerical analysis are performed for wide range of system
parameters when basic flow has constant shear of velocity and
uniform magnetic field is parallel to the basic flow. The special
features consist of magnetohydrodynamic wave mutual transformation
and over-reflection phenomena. The transformation takes place for
arbitrary shear rates and involves all magnetohydrodynamic wave
modes. While the over-reflection occurs only for slow magnetosonic
and Alfv\'en waves at high shear rates. Studied phenomena should
be decisive in the elaboration of the self-sustaining model of
magnetohydrodynamic turbulence in the shear flows.

\end{abstract}

\maketitle

\section{Introduction}               

Flows with inhomogeneous velocity profiles are one of the
prevalent and still not completely understood macro-systems. They
occur in laboratory experiments, industrial applications, the
earth's atmosphere, oceans and many astrophysical objects. We
concentrate attention on smooth shear flows (flows without
inflection point in the velocity profile) that are linearly stable
according to the canonical hydrodynamics. Faults in the
comprehension of shear flow dynamics is caused by the
incompleteness of the classical method of normal modes.
Specifically, method of modes leaves out of account the
non-normality of the linear operators that often produces powerful
transient development of a subset of perturbations
\cite{GL65,CD90,RH93,CRT96,TVY01,VMY01}. Indeed, in the case of
the non-normality of linear operators, corresponding
eigenfunctions are not orthogonal and strongly interfere
\cite{RH93}. Hence, a correct approach should fully analyze the
eigenfunction interference, which is feasible in asymptotic: in
fact, in modal analysis only the asymptotic stability of flow is
studied, while {\it no attention} is paid to any particular
initial value or finite time period of the dynamics -- this period
of the evolution is thought to have no significance and  is left
to speculation.

Recognition of the importance of the non-normality in the linear
stability resulted impressive progress in the understanding of the
shear flow phenomena in 90s of the last century. The early
transient period for the perturbations has been shown to reveal
rich and complicated behavior leading to various consequences. It
has been grasped phenomena that are overlooked in the framework of
the modal analysis.

The progress in the understanding of the shear flow phenomena has
been achieved using so-called non-modal approach (see, e.g., Refs.
\cite{GL65,CRT96,TVY01}), which implicates a change of independent
variables from a laboratory to a moving frame and the study of
temporal evolution of {\it spatial Fourier harmonics} (SFHs) of
perturbations without any spectral expansion in time. Strictly
speaking the non-modal approach is applicable to the flows with
constant velocity shear. However, it is obvious that the results
obtained in the framework of the approach are valid for any shear
flows without inflection point, if the wavelength of modes is
appreciably shorter than the characteristic length scale of the
inhomogeneity.

The progress involves novel results on the time evolution of
vortex and wave perturbations; elaboration of a concept of
self-sustaining turbulence in the spectrally stable shear flows
(that was labelled as bypass concept)
\cite{BDT95,CCHL02,C02,GG94,HR94}. Among the novelties one should
stress (in the context of the present paper) disclosure of the
linear mechanisms of resonance transformation of waves at low
shear rates \cite{CRT96,CCLT97,RPM00} and non-resonant conversion
of vortices to waves at moderate and high shear rates
\cite{CTBM97}. The later phenomena should be inherent to flow
systems. This was shown in papers \cite{RPM00,RPH99}, where
appearances of some waves in solar wind and galactic disks are
explained by the linear wave transformations in the
magnetohydrodynamic (MHD) flows.

Aim of the present work is quantitative and qualitative study of
the linear evolution of three dimensional MHD waves at low and
high shear rates. Equilibrium density, pressure and magnetic field
are assumed to be homogeneous. Specifically, we perform analytical
and numerical study of the important effect caused by linear
forces -- mutual transformations of MHD waves. The waves are
coupled (during a limited time interval) at the specific system
parameters and mutual transformations of two, or even all three of
them occur. Analytical expressions of transformation coefficients
are obtained for wide range of the system parameters. We study the
wave over-reflection phenomena at high shear rates: MHD waves
extract flow energy, are amplified non-exponentially and are
over-reflected.

Mathematical methods used in this paper are similar to ones that
were first developed for quantum mechanical problems: non-elastic
atomic collisions \cite{S32} and non-adiabatic transitions in two
level quantum systems \cite{Z32,L32}. Later, the same asymptotic
methods were successfully applied to various problems (see, e.g.,
Ref. \cite{ZM65}) including interaction of plasma waves in the
media with inhomogeneous magnetic field and/or density \cite{S}.

The paper is organized as follows: employed mathematical formalism
is presented in Sec. II. General properties of resonant
transformation of  wave modes that takes place at small shear
rates, are presented in Sec. III. Particular cases of the resonant
transformation of MHD wave modes are described in Sec. IV. In Sec.
V dynamics of SFHs of the wave modes at high shear rates and
over-reflection phenomenon is studied. Conclusions are given in
Sec. VI.

\section{Mathematical Formalism}

Consider compressible unbounded shear flow with constant shear
parameter (${\bf U}_0(A y,0,0)$) and uniform density $(\rho_0)$,
pressure $(P_0)$ and magnetic field directed along the streamlines
$(\bf B_0 || \bf U_0)$. Linearized ideal MHD equations governing
the evolution of density $(\rho^{\prime})$, pressure
$(p^{\prime})$, velocity $(\bf{u^{\prime}})$ and magnetic field
$(\bf {b^{\prime}})$ perturbations in the flow are:
\begin{equation}
(\partial_t + {\bf U}_0 \cdot {\bf \nabla} )\rho^{\prime} +
{\rho_0} {\bf \nabla} \cdot {\bf{u^{\prime}}} = 0, \label{eq:p1}
\end{equation}
\begin{equation}
(\partial_t + {\bf U}_0 \cdot {\bf \nabla} ){\bf{u^{\prime}}} +
({\bf u^{\prime}} \cdot {\bf \nabla}) {\bf U}_0 + \frac{1}{\rho_0}
{\bf \nabla} p^{\prime} = \frac{-1}{4\pi \rho_0} {\bf B}_0 \times
({\bf \nabla} \times {\bf{b^{\prime}}}),
\end{equation}
\begin{equation}
(\partial_t + {\bf U}_0 \cdot {\bf \nabla}){\bf{b^{\prime}}} -
({\bf B}_0 \cdot {\bf \nabla} ){\bf{u^{\prime}}} + {\bf B}_0 ({\bf
\nabla} \cdot {\bf{u^{\prime}}})= 0,
\end{equation}
\begin{equation}
{\bf \nabla} \cdot {\bf{b^{\prime}}} = 0. \label{eq:p2}
\end{equation}
Further we use the standard technique of the non-modal approach
\cite{GL65}, i.e., expand the perturbed quantities as:
\begin{equation}
\psi^\prime({\bf x},t) = \psi^\prime ({\bf k},t) \exp \left[
i(k_xx + k_y(t)y+k_zz) \right], \label{eq:fou}
\end{equation}
where $\psi^\prime \equiv ({u_x}^{\prime}, {u_y}^{\prime},
{u_z}^{\prime}, \rho^\prime, p^\prime,
{b_x}^{\prime},{b_y}^{\prime}, {b_z}^{\prime})$, $ k_y(t) = k_y(0)
- k_x A t$, and ($k_x$, $k_y(0)$, $k_z$) are the wave numbers of
SFHs at the initial moment of time. It follows from Eq.
(\ref{eq:fou}), that $k_y(t)$ varies in time. This fact can be
interpreted as a "drift" of SFH in phase space. This circumstance
is caused by the fact, that perturbations cannot have a simple
plane wave form in the shear flow due to the shearing background
\cite{CD90}.

Using the following thermodynamic relation  $p^\prime = {c_s}^2
\rho^\prime$, where $c_s$ is sound speed, and introducing
non-dimensional parameters and variables:
\begin{eqnarray}
\begin{array}{ll}
S \equiv A/(V_A k_x)$, ~~$\tau \equiv k_x V_At,~~\beta \equiv
c_s^2/V_A^2,\\ K_z \equiv k_z/k_x,~~K_y(\tau) \equiv k_y/k_x-S
\tau,\\ K^2(\tau)=1+K_y^2(\tau)+K_z^2,~~\rho({\bf k},\tau)\equiv i
{\rho^{\prime}({\bf k},\tau)}/{\rho_0},\\ {\bf b}({\bf k},\tau)
\equiv i{{\bf b}^{\prime}({\bf k},\tau)}/{B_0},~~{\bf v}({\bf
k},\tau) \equiv {\bf u}^{\prime}({\bf k},\tau)/V_A
\end{array}\label{eq:nondim}
\end{eqnarray}
(where $ V_A \equiv B_0/ \sqrt{4 \pi \rho_0}$ is the Alfv\'en
velocity) the set of Eqs. (\ref{eq:p1})-(\ref{eq:p2}) is reduced
to:
\begin{equation}
\frac{{\rm d} \rho}{{\rm d} \tau} = v_x + K_y(\tau) v_y + K_z v_z,
\label{eq:o1}
\end{equation}
\begin{equation} \frac{{\rm d} v_x}{{\rm d} \tau} = - S v_y
- \beta \rho,
\end{equation}
\begin{equation}
\frac{{\rm d} v_y}{{\rm d} \tau} = - \beta K_y(\tau) \rho + (1 +
K_y^2 (\tau)) b_y + K_y(\tau) K_z b_z,
\end{equation}
\begin{equation}
\frac{{\rm d} v_z}{{\rm d} \tau} = - \beta K_z \rho + (1 + K_z^2)
b_z + K_y(\tau) K_z b_y,
\end{equation}
\begin{equation} \frac{{\rm d} b_y}{{\rm d} \tau} = -
v_y,
\end{equation}
\begin{equation} \frac{{\rm d} b_z}{{\rm d} \tau} = -
v_z,
\end{equation}
\begin{equation} b_x+K_y(\tau) b_y + K_z b_z=0.
\label{eq:o2}
\end{equation}

SFH energy density in the non-dimensional form may be defined as a
sum of non-dimensional kinetic, magnetic and compression energy
densities:
\begin{equation}
E({\bf k},\tau) \equiv E^{k}({\bf k},\tau) + E^{m}({\bf k},\tau) +
E^{c}({\bf k},\tau), \label{eq:E}
\end{equation}
where:
\begin{equation}
E^{k}({\bf k},\tau) \equiv \sum_{i=1}^3 |v_i|^2/2, \label{eq:Ek}
\end{equation}
\begin{equation}
E^{m}({\bf k},\tau) \equiv \sum_{i=1}^3 |b_i|^2/2, \label{eq:Em}
\end{equation}
\begin{equation}
E^{c}({\bf k},\tau) \equiv \beta |\rho|^2 /2, \label{eq:Ec}
\end{equation}

For further analysis it is convenient to introduce a new variable
$d \equiv \rho + K_y(\tau)b_y + K_zb_z$ and rewrite the set of
Eqs. (\ref{eq:o1})-(\ref{eq:o2}) in the following form
\cite{RPM00}:
\begin{equation}
 \frac{{\rm d^2}d}{{\rm d}\tau^2} + C_{11}d +
 C_{12}(\tau)b_y+C_{13}b_z = 0,
 \label{eq:o3}
\end{equation}
\begin{equation}
\frac{{\rm d^2} b_y}{{\rm d} \tau^2} + C_{22} (\tau) b_y +
C_{21}(\tau)d+C_{23}(\tau)b_z=0, \label{eq:o4}
\end{equation}
\begin{equation}
\frac{{\rm d^2} b_z}{{\rm d} \tau^2} + C_{33} b_z +
C_{31}d+C_{32}(\tau)b_y=0, \label{eq:o5}
\end{equation}
where: $ C_{11} \equiv \beta$,~~ $ C_{22}(\tau) \equiv
1+(1+\beta)K_y^2(\tau)$, ~~$ C_{33} \equiv 1+(1+\beta)K_z^2$,
~~$C_{12}(\tau) = C_{21}(\tau) \equiv - \beta K_y(\tau)
$,~~$C_{13} = C_{31} \equiv -\beta K_z $ and $C_{23}(\tau) =
C_{32}(\tau) \equiv (1+\beta) K_y(\tau) K_z$.

Combining Eqs. (\ref{eq:o3})-(\ref{eq:o5}) it is easy to show that
the linear evolution of the perturbations has the following
invariant:
\begin{equation}
d^\ast \frac{{\rm d} d}{{\rm d} \tau} - d \frac{{\rm d}
d^\ast}{{\rm d} \tau}+b_y^\ast \frac{{\rm d} b_y}{{\rm d} \tau} -
b_y \frac{{\rm d} b_y^\ast}{{\rm d} \tau} +b_z^\ast \frac{{\rm d}
b_z}{{\rm d} \tau} - b_z \frac{{\rm d} b_z^\ast}{{\rm d}
\tau}=inv, \label{eq:inv}
\end{equation}
Here and hereafter asterisk denotes complex conjugated value. Eq.
(\ref{eq:inv}) represents conservation of wave action of the
system \cite{W65,G84,NS92}.

Eqs. (\ref{eq:o3})-(\ref{eq:o5}) describe the linear dynamics of
MHD waves in the constant shear flow. However, variables $d, b_y,
b_z$ are not normal. This circumstance complicates physical
treatment of perturbation dynamics. Therefore, in the next
subsection normal variables are introduced in the shearless limit.
Afterwards Eqs. (\ref{eq:o3})-(\ref{eq:o5}) are rewritten in the
normal variables and general features of perturbation dynamics as
well as particular characteristics of the system evolution for
different values of parameters $\beta, S$ and $K_z$ are studied in
detail.

\subsection{Shearless limit}

For the sake of clearness of further analysis first we shortly
discuss the shearless limit. Introducing normal variables $
\Psi_i, ~(i=1,2,3)$ \cite{CG}:
\begin{equation}
\Psi_i = \sum_j Q_{ij}\psi_j, \label{eq:nor1}
\end{equation}
where
\begin{eqnarray}
{\bf \psi} = \left(
\begin{array}{cc}
d \\ b_y \\ b_z
\end{array} \right),
\end{eqnarray}
\begin{equation}
{\bf Q} = \left( \begin{array}{rrr} \cos \alpha & \sin \alpha \cos
\gamma & \sin \alpha \sin \gamma \\ \sin \alpha & -\cos \alpha
\cos \gamma & -\cos \alpha \sin \gamma \\ 0 & \sin \gamma & -\cos
\gamma
\end{array} \right).
\label{eq:mx1}
\end{equation}
$\alpha $ and  $\gamma $ stand for Euler angles in the ${\bf
\psi}$-space. After quite long but straightforward algebra it can
be shown, that:
\begin{equation}
\alpha = \arctan \left(\frac{\beta K_\perp}{\Omega_s^2-\beta}
\right), \label{eq:alp0}
\end{equation}
\begin{equation}
\gamma = \arctan \left(\frac{K_z}{K_y} \right), \label{eq:gam0}
\end{equation}
with $ K_\perp^2 \equiv K_y^2+K_z^2$.

In the normal variables Eqs. (\ref{eq:o3})-(\ref{eq:o5}) are
decoupled and reduced to
\begin{equation}
\frac{{\rm d^2} \Psi_i}{{\rm d} \tau^2} + \Omega_i^2 \Psi_i=0.
\label{eq:o7}
\end{equation}
These equations describe independent oscillations with  so-called
fundamental frequencies $\Omega_i$:
\begin{displaymath}
{\Omega^2_{1,2}}=\frac{1}{2}(1+\beta) K^2 \left[1 \pm
\sqrt{1-\frac{4\beta}{(1+\beta)^2 K^2}} ~\right],
\end{displaymath}
\begin{equation}
{\Omega^2_3}=1, \label{eq:fre}
\end{equation}
that are eigen-frequencies of the set of Eqs.
(\ref{eq:o3})-(\ref{eq:o5}) in the shearless limit and can be
easily identified as fast and slow magnetosonic (FMW and SMW) and
Alfv\'en waves (AW), respectively. Hereafter parallel with
notations of eigen-frequencies $\Omega_i$ and eigenfunctions
$\Psi_i$ where $i\equiv(1,2,3)$ we also use the notations
associated with corresponding MHD wave modes, i.e., hereafter
indexes of $\Omega_i$ and $\Psi_i$ vary over $(1,2,3)$ or
equivalently $(F,S,A)$.

Substituting Eqs. (\ref{eq:mx1})-(\ref{eq:gam0}) into
(\ref{eq:nor1}) one can get the following expressions for the
normal variables:
\begin{equation}
\Psi_1\equiv\Psi_F = \frac{(\Omega_S^2 - \beta)\rho + \Omega_S^2
(K_y b_y + K_z b_z )}{\sqrt{(\Omega_S^2 - \beta)^2+\beta^2
K_\perp^2}}, \label{eq:n2}
\end{equation}
\begin{equation}
\Psi_2\equiv\Psi_S = \frac{\beta {K_\perp^2} \rho + (\Omega_S^2 -
\beta K^2 )(K_y b_y + K_z b_z )}{K_\perp \sqrt{(\Omega_S^2 -
\beta)^2+\beta^2 K_\perp^2}}, \label{eq:n3}
\end{equation}
\begin{equation}
\Psi_3\equiv\Psi_A = \frac{K_z b_y - K_y b_z}{\sqrt{K_\perp^2}}.
\label{eq:n4}
\end{equation}

\subsection{Dynamical equations in normal variables}

In shear flows, coefficients in Eqs. (\ref{eq:o3})-(\ref{eq:o5})
vary in time. Therefore the equations governing the dynamics of
SFHs become coupled [in contrast to the shearless equations
(\ref{eq:o7})] and take the following form in the normal
variables:
\begin{equation}
\frac{{\rm d^2} \Psi_i}{{\rm d} \tau^2} + \left( \Omega_i^2 -
\Theta_i \right) \Psi_i =   - \Upsilon_{ik} \Psi_k -
\Lambda_{ik}\frac{{\rm d} \Psi_k}{{\rm d} \tau}~,\label{eq:o8}
\end{equation}
where:
\begin{equation}
\Theta_i =\sum_{j}  {\dot{Q}}_{ij}^2~, \label{eq:aa1}
\end{equation}
\begin{eqnarray}
\Upsilon_{ik} = \left\{
\begin{array}{cc}
\sum_{j} Q_{ij} {\ddot{Q}}_{kj} & i \neq k
\\0 & i = k
\end{array}~, \right.
\end{eqnarray}
\begin{eqnarray}
\Lambda_{ik} = \left\{
\begin{array}{cc}
\sum_{j} 2 Q_{ij} {\dot{Q}}_{kj} & i \neq k
\\0 & i = k
\end{array}~. \right. \label{eq:aa2}
\end{eqnarray}
Here and hereafter overdot denotes $\tau$ derivative. $\Omega_i$
and $Q_{ij}$ are defined by the same expressions as in the
previous subsection [see Eqs. (\ref{eq:mx1})-(\ref{eq:gam0}) and
(\ref{eq:fre})], but with time dependent normalized wave number
$K_y(\tau)=K_y-S\tau$. Expressions for the coefficients in Eq.
(\ref{eq:o8}) are given in the Appendix A.

There are three main differences between Eqs. (\ref{eq:o8}) and
shearless equations (\ref{eq:o7}) caused by the time dependence of
$Q_{ij}$ in the shear flow. These differences correspond to three
channels of energy exchange processes and are responsible for
novel features of linear dynamics of the perturbations in the
shear flows:\\ -- time dependence of the eigen-frequencies
[$\Omega_i=\Omega_i(K_y(\tau))$] causes adiabatic energy exchange
between main flow and perturbations;\\ -- terms on the right hand
side of Eq. (\ref{eq:o8}) describe the coupling between different wave modes;\\
-- additional terms ($\Theta_i$) on the left hand side of Eq.
(\ref{eq:o8}) describe shear induced modification of frequencies.
Influence of these terms become remarkable at high shear rates
($S\gtrsim 1$) and, as it will be described in Sec. V, they are
responsible for the over-reflection phenomenon of MHD wave modes.

Our further efforts are focused on the analysis of this novelties
of the perturbation dynamics.

\subsection{Adiabatic evolution of SFHs}

There are two necessary conditions that should be satisfied for
the validity of WKB approximation. Firstly, shear modified
frequencies $\bar{\Omega}_i^2 \equiv \Omega_i^2-\Theta_i$ must be
slowly varying functions:
\begin{equation}
\dot{\bar{\Omega}}_i \ll \bar{\Omega}_i^2. \label{eq:wkbc}
\end{equation}

Second condition implies that coupling terms in Eq. (\ref{eq:o8})
have to be negligible.

Let us first consider the limit $S\ll 1$. In this case condition
(\ref{eq:wkbc}) reduces to
\begin{equation}
\dot{\Omega}_i \ll {\Omega}_i^2. \label{eq:wkbc1}
\end{equation}
This condition indicates that WKB approximation fails in some
vicinities of the turning points where $\Omega_i(\tau)=0$. But
direct evaluation of Eqs.(\ref{eq:fre}) indicates that none of the
turning points are located near the real $\tau$-axis and therefore
the condition (\ref{eq:wkbc1}) is satisfied for SFHs of all wave
modes at any moment of time for arbitrary values of the parameters
$K_y(\tau),~K_z$ and $\beta$.

If the frequencies of two oscillating modes became equal at some
moment of time, ${\bf Q}$ becomes degenerated. Equivalently, as it
can be seen from direct evaluation of Eqs.
(\ref{eq:aa4})-(\ref{eq:aa5}), coupling coefficients of
corresponding oscillations in Eq. (\ref{eq:o8}) which otherwise
are of order $S$ or $ S^2$ become infinity. So, at $S \ll 1$
evolution of SFHs of the wave modes is adiabatic except some
vicinities of resonant points where
$\Omega_i(\tau)=\Omega_j(\tau)$, or equivalently during some time
intervals when frequencies of different wave modes are close to
each other. Analysis of Eq. (\ref{eq:fre}) yields that all the
resonant points are located on the axis ${\rm Re}[K_y(\tau)]=0$ in
the complex $\tau$-plane. Consequently, these time intervals can
appear only in a certain $\Delta K_y$ vicinity of the point
$K_y(\tau)=0$ (exact conditions for effective coupling between
different modes will be formulated in the next section).

Now consider the case of moderate and high shear rates when
condition $S\ll 1$ is not satisfied. Again, analysis of turning
points where $\bar{\Omega}_i(\tau)=0$ and resonant points where
$\bar{\Omega}_i(\tau)=\bar{\Omega}_j(\tau)$ is important. But in
contrast with the limit $S\ll 1$, there is no small parameter in
the problem. Therefore, in general WKB approximation may fail
during the whole evolution. Combining Eqs. (\ref{eq:fre}) and
(\ref{eq:alp0})-(\ref{eq:gam0}) after long but straightforward
calculations one can conclude, that both conditions for the
validity of WKB approximation formulated above hold for SFHs of
all MHD wave modes for arbitrary $\beta$ and $K_z$ at least for
high values of $|K_y(\tau)|$:
\begin{equation}
|K_y(\tau)| \gg S.\label{eq:cos}
\end{equation}
This circumstance is crucially important for the analysis of the
mode coupling at high shear rates presented in Sec. V.

If the conditions for the validity of WKB approximation holds,
temporal evolution of SFHs can be described by standard WKB
solutions:
\begin{equation}
\Psi_i^\pm = \frac{D_i^\pm}{\sqrt {\Omega_i(\tau)}}e^{\pm i \int
\Omega_i(\tau)d\tau}, \label{eq:wkb}
\end{equation}
where $D_i^\pm$ are WKB amplitudes of the wave modes with positive
and negative phase velocity along $X$-axis respectively. All the
physical quantities can be easily found by combining Eqs.
(\ref{eq:n2})-(\ref{eq:n4}). Combining these equations after long
but straightforward algebra one can check, that the energies of
the wave modes satisfy standard relations of adiabatic evolution:
\begin{equation}
E_i = \Omega_i(\tau) (|{D_i^+}|^2 + |{D_i^-}|^2), \label{eq:Ewk}
\end{equation}
so asymptotically ($\tau \rightarrow \infty $) total energy of SFH
of SMW and AW become constant whereas the total energy of SFH of
FMW increases linearly with $\tau$. In other words, SFH of FMW can
effectively extract energy from the basic flow. It follows from
Eq. (\ref{eq:Ewk}), that  $ |{D_i^\pm}|^2$ can be interpreted as
number of wave particles (so called plasmons) in analogy with
quantum mechanics.

Using Eq. (\ref{eq:wkb}) we can reduce Eq. (\ref{eq:inv}) to the
following form:
\begin{equation}
\sum_i |{D_i^+}|^2 - \sum_i |{D_i^-}|^2 = inv. \label{eq:in1}
\end{equation}
This equation is crucially important for the study of
non-adiabatic processes in the considered flow: as it was
mentioned above, the evolution of SFHs of the wave modes is always
adiabatic for arbitrary $K_y(\tau)$ which is outside some $\Delta
K_y$ vicinity of the point $K_y(\tau_+)=0$, i.e.:
\begin{equation}
|K_y(\tau)| > \Delta K_y, \label{eq:in111}
\end{equation}
where the value of $\Delta K_y$ depends on the specific parameters
of the problem. From the point of view of temporal evolution this
means that non-adiabatic processes can be important only during
the time interval $\Delta \tau \equiv \Delta K_y/S$ in the
vicinity of $\tau_+$, where condition (\ref{eq:in111}) fails.

Consequently, the dynamics of SFHs is the following: assume at the
initial moment of time $K_y(0) > \Delta K_y$. According to Eq.
(\ref{eq:in111}) the dynamics of SFHs is adiabatic initially. Due
to the linear drift in the ${\bf k}$-space, $K_y(\tau)$ decreases
and when condition (\ref{eq:in111}) fails, the dynamics of SFHs
become non-adiabatic. Duration of non-adiabatic evolution is
$\Delta \tau$. Afterwards, when $K_y(\tau) < -\Delta K_y$, the
evolution of SFHs becomes adiabatic again.

Denote WKB amplitudes of  SFHs of the wave modes on the {\it left
and right sides} of the area of non-adiabatic evolution (i.e., for
$\tau<\tau_+-\Delta \tau/2$ and $\tau
 > \tau_+ + \Delta \tau/2$) by ${D_{i,L}^\pm}$ and ${D_{i,R}^\pm}$
respectively. In other words, ${D_{i,L}^\pm}$ and ${D_{i,R}^\pm}$
are WKB amplitudes before and after non-adiabatic evolution. Eq.
(\ref{eq:in1}) provides important relation between WKB amplitudes
on the left and right sides of the area of non-adiabatic
evolution, independent of the behaviour/dynamics of the system in
the non-adiabatic area:
\begin{equation}
\sum_i |D_{i,L}^+|^2 - \sum_i |D_{i,L}^-|^2 = \sum_i |D_{i,R}^+|^2
- \sum_i |D_{i,R}^-|^2. \label{eq:in11}
\end{equation}

\section{general properties of mode coupling}

As it was mentioned above WKB approximation is valid for arbitrary
$K_y(\tau)$ except some vicinity of the point $K_y(\tau_+)$. In
the formal analogy with S-matrix of the scattering theory \cite{K}
and transition matrix from the theory of multi-level quantum
systems \cite{F,LL}, one can connect ${D_{i,R}^\pm}$ with
${D_{i,L}^\pm}$ by $6\times6$ transition matrix:
\begin{equation}
\left( \begin{array}{cc} {\bf D}^+_{R} \\ {\bf D}^-_{R}
\end{array} \right) = \left( \begin{array}{cc} {\bf T}^{++} & {\bf T}^{+-}  \\ {\bf T}^{-+} & {\bf T}^{--}
\end{array} \right) \left( \begin{array}{cc} {\bf D}^+_{L} \\ {\bf D}^-_{L}
\end{array} \right),
\label{eq:mxt}
\end{equation}
where ${\bf D}^{\pm}_{L}$ and ${\bf D}^{\pm}_{R}$ are $1\times3$
matrices and ${\bf T}^{\pm\pm}$ are $3\times3$ matrices.

Due to the fact that all components of the matrix ${\bf C}$ in the
set of Eqs. (\ref{eq:o3})-(\ref{eq:o5}) are real and
$C_{ij}=C_{ji}$ \cite{F}:
\begin{equation}
T_{ij}^{++}=[T_{ij}^{--}]^\ast \equiv T_{ij}, \label{eq:t1}
\end{equation}
\begin{equation}
T_{ij}^{+-}=[T_{ij}^{-+}]^\ast \equiv \overline{T}_{ij}.
\label{eq:t2}
\end{equation}

Substituting Eq. (\ref{eq:mxt}) in Eq. (\ref{eq:in1}) one can get:
\begin{equation}
\sum_j |T_{ij}|^2 - \sum_j |\overline {T}_{ij}|^{2} =1
\label{eq:in2}
\end{equation}

In general, the components of transition matrix in Eq.
(\ref{eq:mxt}) are complex. This means, that the interaction of
different wave modes changes not only the absolute values of
$D_i^\pm $, but also their phases. We call the absolute value of
the transition matrix components $|T_{ij}|$ and $|\overline
T_{ij}|$ the transformation coefficients of corresponding wave
modes:\\ -- $|T_{ij}|$ represents the transformation coefficient
of $j$ to $i$ mode, that has the same sign of the phase velocity
along $X$-axis (i.e., transmitted mode $i$);\\ -- $|\overline
T_{ij}|$ represents the transformation coefficient of $j$ to $i$
mode, that has the opposite sign of the phase velocity along
$X$-axis (i.e., reflected mode $i$).

It has to be noted that in the similar problems of quantum
mechanics \cite{LL,F}, $|T_{ij}|^{2}$ and $|\overline
{T}_{ij}|^{2}$ represent transitions probabilities between
different quantum states.

As it follows from Eqs. (\ref{eq:Ewk}) and (\ref{eq:mxt}), if
initially only one, for instance $j$ mode exists, i.e.,
$D_{i,L}^\pm \equiv 0$ for $i \neq j$, the energies of transformed
waves do not depend on the phases of transition matrix elements
and are entirely determined by $|T_{ij}|$ and $|\overline
T_{ij}|$. In the presented paper we concentrate attention on
transformation coefficients and do not regard the problem of the
phase multipliers of transition matrix elements.

Now consider the case $S \ll 1$, for which the condition
(\ref{eq:wkbc}) holds at arbitrary $\tau$. It is well known
\cite{F,LL}, that in this case components of ${\bf \overline {T}}$
are exponentially small with respect to the large parameter $1/S$
and can be neglected. Consequently, Eq. (\ref{eq:mxt}) decomposes
and reduces to:
\begin{eqnarray}
&& {\bf D}^+_{R} = {\bf T} {\bf D}^+_{L}, \\ && {\bf D}^-_{R} =
{\bf T}^{\ast} {\bf D}^-_{L}. \label{eq:mxt01}
\end{eqnarray}
From physical point of view this means, that only wave modes with
the same sign of the phase velocity along $X$-axis can effectively
interact, i.e., the wave reflection is negligible.

Eq. (\ref{eq:in2}) reduces to the unitary condition for ${\bf T}$:
\begin{equation}
\sum_j |T_{ij}|^2 =1. \label{eq:ort2}
\end{equation}

Due to the fact that condition (\ref{eq:wkbc}) is satisfied, the
only reason for the failure of WKB approximation could be
closeness of the wave mode frequencies, i.e., closeness of at
least one resonant point (point where two of more fundamental
frequencies become equal) to the real $\tau$-axis \cite{F,LL}. In
the later case coupling of wave modes becomes effective. Analysis
of Eq. (\ref{eq:fre}) shows that all the resonant points are
located on the axis ${\rm Re}[K_y(\tau)]=0$ in the complex
$\tau$-plane. That is why the effective transformation of wave
modes can take place only in the vicinity of the moment of time
$\tau_+$ where $K_y(\tau_+)=0$.

First of all let us discuss some general properties of wave
resonant interaction (mathematical details are discussed in the
Appendix B):

(i) for effective coupling between different (for instance $i$ and
$j$) wave modes there should exist a time interval (so-called
resonant interval) where \cite{KS,LL}:
\begin{equation}
|\Omega_i^2-\Omega_j^2| \lesssim |\Lambda_{ij}\Omega_i|.
\label{eq:co1}
\end{equation}
If this condition is not satisfied transformation coefficients are
exponentially small with respect to the large parameter $ 1/S$,
namely \cite{F,LL}:
\begin{equation}
T_{ij} \sim  \exp\left( - \left|  {\rm Im}
\int_{\tau_0}^{\tau_{ij}}(\Omega_i-\Omega_j)d\tau \right| \right).
\label{eq:tr2}
\end{equation}
Here and hereafter the signs of absolute magnitude for
transformation coefficients are omitted, i.e., under $T_{ij}$ we
mean $|T_{ij}|$. In Eq. (\ref{eq:tr2}), $ \tau_0$ is arbitrary
point on the real $ \tau$-axis and $\tau_{ij}$ is the nearest to
the real $ \tau$-axis resonant point where $\Omega_i(\tau_{ij}) =
\Omega_j(\tau_{ij})$.

The characteristic equation of the set (\ref{eq:o3})-(\ref{eq:o5})
is real and symmetric with respect to the transform $K_y(\tau)
\rightarrow -K_y(\tau) $. Therefore, resonant points always appear
as complex conjugated pairs: if $\tau_{ij}$ is a resonant point,
so is its complex conjugated one $\tau_{ij}^\ast$.

(ii) the following equation holds for resonant interaction of two
wave modes,  e.g., i and j (i.e., when condition (\ref{eq:co1}) is
satisfied only for two wave modes):
\begin{equation} T_{ij}=T_{ji}. \label{eq:tr3}
\end{equation}
This symmetry property, which follows from unitarity property
(\ref{eq:ort2}), holds for resonant interaction of two wave modes
only. If at the same interval of time there is effective coupling
of more then two wave modes, then Eq. (\ref{eq:tr3}) fails.

(iii) if in the neighborhood of the real $\tau$-axis only a pair
of complex conjugated first order resonant points $\tau_{ij}$ and
$\tau_{ij}^\ast$ exists [the resonant point is called of order $n$
if $(\Omega_i^2-\Omega_j^2) \sim (\tau - \tau_{ij})^{n/2}$ in the
neighborhood of $\tau_{ij}$], the transformation coefficients are
\cite{LL,ZM65}:
\begin{equation}
T_{ij} = \exp\left( - \left| {\rm Im}
\int_{\tau_0}^{\tau_{ij}}(\Omega_i-\Omega_j)d\tau \right| \right)
[1+ O(S^{1/2})]. \label{eq:tr4}
\end{equation}
There are two remarks about this equation. Firstly, it shows, that
only dispersion equation is needed to derive the transformation
coefficient with accuracy $S^{1/2}$ in the case of the first order
resonant points. In other words, only the solution of the
characteristic equation of the governing set of Eqs.
(\ref{eq:o1})-(\ref{eq:o2}) is needed to derive the transformation
coefficient. Secondly, Eq. (\ref{eq:tr4}) is valid also in the
case of strong wave interaction. For example, if complex
conjugated resonant point of the first order tends to the real
$\tau$-axis, then $ T_{ij}\rightarrow 1$. According to Eq.
(\ref{eq:ort2}), this means that one wave mode is fully
transformed into another.

(iv) for the second or higher order resonant points analytical
expression of transformation coefficient can be derived only in
the case of weak interaction ($ T_{ij}\ll 1,~~ i \neq j$). Namely,
if there exist a pair of complex conjugated second order resonant
points and condition (\ref{eq:co1}) is not satisfied,
transformation coefficients are:
\begin{equation}
T_{ij} = \frac{\pi}{2} \exp\left(- \left| {\rm Im}
\int_{\tau_0}^{\tau_{ij}}(\Omega_i-\Omega_j)d\tau \right| \right)
[1+ O(S^{1/2})]. \label{eq:tr5}
\end{equation}

Let us note once again, that this equation is valid only for
$T_{ij} \ll 1$. Derivation of this formula is presented in
Appendix B.

In earlier studies, an attention was always paid to the resonant
points of the first order \cite{LL,F,ZM65}. As it will be shown
later, all the resonant points are of the second order in the case
of resonant coupling between SFHs of AW and the magneto-sonic wave
modes. This is not an unique property of the evolution of MHD wave
modes. It can be readily shown that these type of resonances
naturally appear in systems of three or more coupled oscillators.

(v) from Eq. (\ref{eq:co1}) it can be shown (see Appendix B), that
the time scale of the resonant interaction is as follows:
\begin{equation}
\Delta \tau_{ij} \sim S^{-1+1/n}, \label{eq:dt1}
\end{equation}
where $n $ is the order of the resonant point. Whereas the time
scale of adiabatic evolution $\Delta \tau \sim 1/S$. Therefore if
$ S \ll 1$
\begin{equation}
\Delta \tau_{ij} \ll \Delta \tau,\label{eq:dt2}
\end{equation}
i.e., resonant interaction of waves is much faster process then
energy exchange between background flow and wave modes.
Consequently, conservation of wave action (\ref{eq:in2}) reduces
to energy conservation during the resonant interaction of wave
modes.

\section{Specific limits of the resonant transformation of MHD wave modes}

In this section we study specific cases of MHD wave coupling at $S
\ll 1$:

(A) Two dimensional (2D) problem when $K_z \equiv b_z \equiv 0$.
In this case Alfv\'en waves are absent, and obtained set of
equations describes the coupling between FMW and SMW.

(B) $\beta \ll 1$. In this limit $\Omega_{F,A}\gg \Omega_S$ and
only mutual transformation of FMW and AW is possible.

(C) $\beta \gg 1$. In this limit $\Omega_{A,S}\ll \Omega_F$ and
mutual transformation of SMW and AW can be effective.

(D) $\beta \sim 1$. In this case frequencies of all the MHD waves
can be of the same order and mutual transformations of all the
modes is possible.

As it was mentioned above resonant transformations of MHD wave
modes in shear flows are investigated recently
\cite{CRT96,CCLT97,RPM00}. The content of this section is
concerned on detailed quantitative analysis of the problem. In
particular - derivation of analytical expressions of
transformation coefficients.

\subsection{2D problem}

To derive equations in 2D case one has to assume $K_z \equiv b_z
\equiv 0$. This limit excludes Alfv\'en waves and Eqs.
(\ref{eq:o3})-(\ref{eq:o5}) reduce to:
\begin{equation}
\frac{{\rm d^2}d}{{\rm d}\tau^2} + C_{11}d + C_{12}(\tau)b_y= 0,
\label{eq:o31}
\end{equation}
\begin{equation}
\frac{{\rm d^2} b_y}{{\rm d} \tau^2} + C_{22} (\tau) b_y +
C_{21}(\tau)d=0, \label{eq:o41}
\end{equation}
where: $ C_{11} \equiv \beta$,~~ $ C_{22}(\tau) \equiv
1+(1+\beta)K_y^2(\tau)$,~~$C_{12}(\tau) = C_{21}(\tau) \equiv -
\beta K_y(\tau) $. Eqs. (\ref{eq:o31})-(\ref{eq:o41}) describe
coupled evolution of SFHs of FMW and SMW. Frequencies of the modes
are given by Eq. (\ref{eq:fre}) where now $K^2 \equiv
1+K_y^2(\tau)$. Normal variables are defined by Eqs.
(\ref{eq:n2})-(\ref{eq:n3}) with substitution $K_z = b_z = 0$.

Solving the equation $\Omega_F=\Omega_S $ one can easily obtain
that there is only a pair of complex conjugated resonant points of
the first order in the complex $\tau$-plane:
\begin{equation}
K_y(\tau_{FS}) = i \frac{\beta-1}{\beta+1},~~K_y(\tau_{FS}^\ast) =
-i \frac{\beta-1}{\beta+1}. \label{eq:res0}
\end{equation}
Noting, that in the neighborhood of resonant points:
\begin{equation}
\Omega_F - \Omega_S \approx \left({\frac{\beta+1}{2}}\right)^{1/2}
\left [ K_y^2(\tau)+\left( \frac{\beta-1}{\beta+1}\right)^2
\right]^{1/2}, \label{eq:fr2}
\end{equation}
from Eq. (\ref{eq:tr4}) one can easily obtain:
\begin{equation}
T_{FS} \approx \exp \left[ -\frac{\pi \sqrt{1+\beta}}{4\sqrt{2} S}
\left(\frac{\beta-1}{\beta+1}\right)^2 \right]. \label{eq:tr6}
\end{equation}

It is seen from Eq. (\ref{eq:res0}), that if $\beta \rightarrow
1$, the resonant points tends to the real $\tau$-axis. Then from
Eq. (\ref{eq:tr6}) it follows that $T_{FS} \rightarrow 1$.
According to Eq. (\ref{eq:ort2}), it means that one wave mode
totally transforms into another.

Dependence of transformation coefficient on $\beta$ is presented
in Fig. \ref{fig:fig1} for $ S=0.05$ and $S=0.02$. Dotted line
shows transformation coefficient obtained by numerical solution of
the set of Eqs. (\ref{eq:o3})-(\ref{eq:o5}), where initial
conditions are chosen by WKB solutions (\ref{eq:wkb}) far on the
left hand side of resonant time interval ($K_y(0) \ll -1$) and
solid line is the curve of analytical solution (\ref{eq:tr6}).

\subsection{Low $\beta$ regime}

In this case the frequency of SMW is far less then frequencies of
FMW and AW  ($\Omega_S \ll \Omega_F,\Omega_A $). Therefore, the
coupling of SMW with other MHD modes is exponentially small with
respect to the parameter $1/S$ and can be neglected. Consequently,
in the set of Eqs. (\ref{eq:o3})-(\ref{eq:o5}) equation for $d$
decouples and equations for $b_y $ and $b_z$ describes coupled
evolution of SFHs of AW and FMW:
\begin{equation}
\frac{{\rm d^2} b_y}{{\rm d} \tau^2} + \left( 1+ K_y^2(\tau)
\right) b_y = -K_y(\tau) K_z b_z, \label{eq:o14}
\end{equation}
\begin{equation}
\frac{{\rm d^2} b_z}{{\rm d} \tau^2} + \left( 1 + K_z^2 \right)
b_z = -K_y(\tau) K_z b_y, \label{eq:o15}
\end{equation}
Normalized frequencies of the coupled wave modes are:
\begin{equation}
\Omega_F^2(\tau) = 1 + K_z^2 + K_y^2(\tau),~~~\Omega_A^2 =
1.\label{eq:fr3}
\end{equation}
From this equation it follows that there are two complex
conjugated second order resonant points:
\begin{equation}
K_y(\tau_{FA}) = i K_z,~~~K_y(\tau_{FA}^\ast) = -i
K_z.\label{eq:fr311}
\end{equation}
Necessary condition for effective coupling expressed by Eq.
(\ref{eq:co1}) now takes the form:
\begin{equation}
|K_z^3| \leq S. \label{eq:co2}
\end{equation}
Thus, the critical parameter is $\delta \equiv K_z/S^{1/3}$.

Consider the case $\delta \gg 1$. For the calculation of the
transformation coefficients one can use general analysis presented
in Sec. III and Appendix B. Specifically, Eq. (\ref{eq:tr5})
yields:
\begin{equation}
T_{FA} = \frac{\pi}{2} \exp\left( \frac{-|\phi_{FA}(K_z)|}{2S}
\right),\label{eq:tr7}
\end{equation}
where:
\begin{equation}
\phi_{FA}(K_z) = \left( 1+K_z^2 \right) \arctan (K_z)-K_z.
\end{equation}
If in addition $ K_z \ll 1$, then Eq. (\ref{eq:tr7}) reduces to:
\begin{equation}
T_{FA} \approx \frac{\pi}{2} \exp\left( -\frac{\delta^3}{3}
\right).\label{eq:tr8}
\end{equation}

Analytical expression for the transformation coefficients can be
derived also in the opposite limit $ \delta \ll 1$, ($K_z \ll
S^{1/3}$). Taking into account the relation $\Omega_S^2 \approx
\beta$, one can readily obtain from Eqs. (\ref{eq:n2}) and
(\ref{eq:n4}) that $b_y$ and $b_z$ coincide with the
eigenfunctions of SFHs of FMW and AW respectively accurate to the
terms of order $K_z^2$. Consequently, terms on the right hand
sides of Eqs. (\ref{eq:o14})-(\ref{eq:o15}) represent the coupling
terms with accuracy $K_z^2$. Since $K_z \ll S^{1/3}$, coupling is
weak and feedback in the set of Eqs. (\ref{eq:o14})-(\ref{eq:o15})
strongly depends on the amplitudes of the wave modes. Then it
follows that if initially only AW exists, feedback of FMW on AW is
negligible and the set of Eqs. (\ref{eq:o14})-(\ref{eq:o15})
reduces to:
\begin{equation}
\frac{{\rm d^2} \Psi_F}{{\rm d} \tau^2} + \left[ 1+ K_y^2(\tau)
\right] \Psi_F = - K_y(\tau) K_z \Psi_A. \label{eq:o141}
\end{equation}
\begin{equation}
\frac{{\rm d^2} \Psi_A}{{\rm d} \tau^2} + \Psi_A =0.
\label{eq:o142}
\end{equation}

Using the solution of Eq. (\ref{eq:o142}) and well known
expressions for the solution of linear inhomogeneous second order
differential equation, in the considered limit ($\delta \ll 1$),
we obtain:
\begin{equation}
T_{FA} \approx 2^{2/3} \delta \int_{0}^{\infty} x \sin \left(
\frac{x^3}{3}- \frac{\delta^2}{2^{2/3}}x \right) d x.
\label{eq:tr9}
\end{equation}
Note that:
\begin{equation}
\int_{0}^{\infty} x \sin \left( \frac{x^3}{3}- \gamma x \right) d
x \equiv \pi \frac{\partial}{\partial \gamma}Ai(-\gamma)
\end{equation}
and using the expansion of Airy function $Ai(\gamma)$ into power
series \cite{AS} we finally obtain:
\begin{equation}
T_{FA} \approx \frac{2^{2/3}\pi}{3^{1/3}\Gamma \left(
\frac{1}{3}\right)} \delta \left( 1 - \frac{\Gamma\left(
\frac{1}{3}\right)}{2^{7/4} 3^{1/3} \Gamma \left(
\frac{2}{3}\right)} \delta^4 \right). \label{eq:tr10}
\end{equation}

Results of numerical solution of the initial set of equations
(\ref{eq:o1})-(\ref{eq:o2}) (solid line) as well as analytical
expressions (\ref{eq:tr8}) (dash-dotted line) and (\ref{eq:tr10})
(dashed line) are presented in Fig. \ref{fig:fig2}. It shows that
the transformation coefficient reaches its maximal value
$(T_{FA}^2)_{max} = 1/2$ at $\delta^{cr}$ that can be found
numerically or alternatively by finding the maximum of the
analytical expression presented by Eq. (\ref{eq:tr10}):
\begin{equation}
\delta^{cr} = \left( \frac{2^{7/4} 3^{1/3} \Gamma \left(
\frac{2}{3}\right)}{5 \Gamma\left(
\frac{1}{3}\right)}\right)^{1/4}. \label{eq:de1}
\end{equation}

Eq. (\ref{eq:de1}) is in perfect accordance with numerically
calculated $\delta^{cr}$ (see Fig. \ref{fig:fig2}) despite the
failure of Eq. (\ref{eq:tr10}) at $\delta \sim 1$. This fact can
be explained as follows: the only reason of failure of Eq.
(\ref{eq:tr10}) is the neglect of the feedback in Eq.
(\ref{eq:o142}). The feedback changes the value of the
transformation coefficient but does not affect on the value of
$\delta^{cr}$.

$({T_{FA}}^2)_{max} = 1/2$ means that only half of the energy of
FMW can be transformed into AW and vice versa even in the optimal
regime. It has to be noted, that Landau-Zener theory
\cite{L32,Z32} provides the same maximum value for the transition
probability in two-level quantum mechanical systems.

\subsection{High $\beta$ regime}

In this case $\Omega_S,\Omega_A \ll \Omega_F$. Consequently, the
coupling of AW and SMW with FMW are exponentially small with
respect to the parameter $1/S$ and can be neglected.

Analysis of Eq. (\ref{eq:fre}) provides that for the coupling of
AW and SMW there exist two complex conjugated second order
resonant points (solutions of the equation $\Omega_S = \Omega_A$),
that are also given by Eq. (\ref{eq:fr311}). The condition of
effective coupling (\ref{eq:co1}) takes the form:
\begin{equation}
\frac{|K_z|^3}{\sqrt {1+K_z^2}} \leq  \beta S.\label{eq:co4}
\end{equation}
If this condition fails, then Eq. (\ref{eq:tr5}) yields
exponentially small transformation coefficient:
\begin{equation}
T_{SA} = \frac{\pi}{2} \exp\left( \frac{-|\phi_{SA}(K_z)|}{2\beta
S} \right),\label{eq:tr11}
\end{equation}
where
\begin{equation}
\phi_{SA}(K_z) = K_z -\frac {\mbox{arcsinh} \left( K_z \right)
}{\sqrt {1+K_z^2}}.
\end{equation}
If additionally $ K_z \ll 1$, then Eq.(\ref{eq:tr11}) reduces to:
\begin{equation}
T_{SA} \approx \frac{\pi}{2} \exp\left( -\frac{|K_z|^3}{3\beta S}
\right).\label{eq:tr12}
\end{equation}

Consider the transformation process when the condition of
effective coupling Eq. (\ref{eq:co4}) is satisfied. Similar to the
low $ \beta$ regime, in this case it is also possible to derive
the set of two second order coupled equations that describe the
linear coupling of AW and SMW. Expressing the density perturbation
$\rho$ from the condition $\Psi_F \equiv 0$ (thus eliminating the
terms that describe interaction of AW and SMW with FMW) and
combining Eqs. (\ref{eq:o3})-(\ref{eq:o5}) one can obtain:
\begin{equation}
\frac{{\rm d^2} b_y}{{\rm d} \tau^2} + \left( 1 + \theta
K_y^2(\tau) \right) b_y = -\theta K_y(\tau) K_z b_z,\label{eq:o16}
\end{equation}
\begin{equation}
\frac{{\rm d^2} b_z}{{\rm d} \tau^2} + \left( 1 + \theta K_z^2
\right) b_z = -\theta K_y(\tau) K_z b_y,\label{eq:o17}
\end{equation}
where:
\begin{equation}
\theta \equiv  1- \Omega_S^2 \frac{\beta}{\beta - \Omega_S^2}
\approx -\frac{1}{\beta K^2(\tau)}.
\end{equation}
As it follows from Eqs. (\ref{eq:o16})-(\ref{eq:o17}) $T_{AS}
\equiv 0$ at $K_z=0$, i.e., there is no interaction between AW and
SMW.

In contrast with low $ \beta$ limit, there is no unique parameter
in high $\beta$ limit that totally describes the transformation
process. In this limit there are two such parameters $K_z $ and
$\beta S$.

First consider the case $ \beta S \ll 1$. Then the condition
(\ref{eq:co4}) reduces to $\delta_1 \equiv |K_z|/(\beta S)^{1/3}
\leq 1 $. Note, that the sign of $\theta$ does not affect the
transformation coefficient and $K^2(\tau) \approx 1$ in the
resonant area. Thus we conclude that the properties of the wave
transformation is the same as in the case of transformation of AW
and FMW. Namely, if $\delta_1 \ll 1$, then the leading terms of
asymptotic expressions of transformation coefficient is given by
Eq. (\ref{eq:tr10}) with $\delta$ replaced by $\delta_1$:
\begin{equation}
T_{AS} \approx \frac{2^{2/3}\pi}{3^{1/3}\Gamma \left(
\frac{1}{3}\right)} \delta_1 \left( 1 - \frac{\Gamma\left(
\frac{1}{3}\right)}{2^{7/4} 3^{1/3} \Gamma \left(
\frac{2}{3}\right)} \delta_1^4 \right).
\end{equation}

At $ \beta S \ll 1$, the transformation coefficient reaches its
maximum $(T_{AS}^2)_{max} =1/2$ at $\delta_1^{cr}$ that coincides
with $\delta^{cr}$ defined by Eq. (\ref{eq:de1}).

Dependence of the transformation coefficient on $K_z $ obtained by
numerical solution of the initial set of equations
(\ref{eq:o3})-(\ref{eq:o5}) for $\beta S = 0.025 $, $\beta S = 0.5
$ and $\beta S = 1 $ are presented in Fig. \ref{fig:fig3}.

Consider the case when $\beta S$ is not small.  The numerical
study shows that in this case the properties of transformation
process is totally different (see Fig. 3):\\ -- transformation
coefficient $T_{AS}$ does not depend on $\beta S $ at $K_z \ll 1$
\begin{equation}
T_{AS} \approx 2.05 K_z, \label{eq:tr15}
\end{equation}
-- $(T_{AS})_{max}=1$, i.e., total transformation of one wave mode
into another is possible.

\subsection{$\beta \sim 1$ regime}

In the case of $\beta \sim 1$, the frequencies of all MHD wave
modes are of the same order and no simplification of the set of
Eqs. (\ref{eq:o3})-(\ref{eq:o5}) is possible. Analysis of Eq.
(\ref{eq:fre}) shows that there exist the pair of complex
conjugated first order resonant points:
\begin{equation}
{K_y}(\tau_{1,2}) = \pm i \sqrt{ K_z^2+\left(
\frac{\beta-1}{\beta+1} \right)^2}
\end{equation}
and the pair of complex conjugated second order resonant points:
\begin{equation}
{K_y}(\tau_{3,4}) = \pm i K_z.
\end{equation}

No analytical expressions can be obtained for transformation
coefficients if more than two wave modes are effectively coupled.

Numerical study of the set of Eqs. (\ref{eq:o3})-(\ref{eq:o5}) is
performed as follows: WKB solutions (\ref{eq:wkb}) are used to
obtain initial values of $d, b_y, b_z$ and their first derivatives
for different wave modes separately far on the left side of the
resonant interval ($K_y(0) \gg 1$). After passing through the
resonant interval (i.e., for any $\tau$, for which $K_y(\tau) \ll
-1$), WKB solutions were used again to determine the intensities
of the transformed wave modes.

Figure \ref{fig:fig4} shows results obtained by numerical solution
of the set of Eqs. (\ref{eq:o3})-(\ref{eq:o5}), for initial SMW.
Namely, transformation coefficients ($T_{FS}, T_{AS}, T_{SS}$) vs
$K_z$ are presented for different values of $\beta$ and $S$.

According to our numerical study, qualitative character of wave
transformation process is similar to the cases described in the
previous sections. However, there are some differences. The most
interesting is the failure of symmetry property (\ref{eq:tr3}).
The presence of third effectively interacting wave mode leads to
the asymmetry of two wave mode interactions, e.g., intensity of AW
generated by SMW differs from the intensity of the inverse
process.

\section{High shear regime: wave over-reflection}

Let us consider the dynamics of SFHs of MHD wave modes in the flow
with moderate and high shear rates ($S \gtrsim 1 $). In this case
the dynamics is strongly non-adiabatic [$\Theta_i$ terms can not
be neglected in the left hand side of Eqs. (\ref{eq:o8})] except
time intervals when $|K_y(\tau)| \gg S$. The existence of these
adiabatic intervals permits to study wave interaction based on the
asymptotic analysis presented in Sec III.

The main novelty that appears in the flow at high shear rates is
that the dynamics involves wave reflection/over-reflection
phenomena. Specifically, $\overline {T}_{ij}$ can not be neglected
and becomes important in Eq. (\ref{eq:in2}). Physically it means
that initial $\Psi^+_i $ mode can be effectively transformed into
$\Psi^-_j $ modes (wave modes with phase velocity directed
opposite to phase velocity of initial wave mode with respect to
$X$-axis). Thus, at high shear rates, the wave dynamics represents
an interplay of transformation and (over)reflection phenomena that
are described by ${T}_{ij}$ and $\overline {T}_{ij}$.

Dependence of ${T}_{ij}$ and $\overline {T}_{ij}$ on $K_z$
obtained by numerical solution of the initial set of equations
(\ref{eq:o3})-(\ref{eq:o5}) for different values of parameters
$\beta$ and $S$ are presented in Figs.
\ref{fig:fig5}-\ref{fig:fig8}. Figs. \ref{fig:fig5}-\ref{fig:fig7}
show the cases when initially only SMW with positive phase
velocity along $ X$-axis exists, i.e., only $D_{S,L}^+ \neq 0 $.
Whereas in Fig. \ref{fig:fig8} initially only AW with positive
phase velocity along $ X$-axis exists, i.e., only $D_{A,L}^+ \neq
0 $.

In Fig. \ref{fig:fig5}, $\beta = 2$ and $S = 1$ (dashed lines) and
$S = 4$ (solid lines). Both, wave transformation and reflection
phenomena appear and enhance with increase of $S$. Coupling of SMW
with FMW is dominant at $K_z \ll 1$, while coupling of SMW with AW
is dominant at $K_z \gtrsim 1$. Figure \ref{fig:fig6} shows the
same plots for high $\beta$ and moderate $S$. Particularly, dashed
lines correspond to $\beta = 50$ and $S = 1$, whereas solid lines
correspond to $\beta = 50$ and $S = 2$. For this range of
parameters $\beta^{1/2} \gg S$ and according to Eq. (\ref{eq:tr6})
${T}_{FS}, \overline {T}_{FS} \approx 0$. Hence, SMW and FMW are
not coupled. Fig. \ref{fig:fig6} shows one more new feature of the
wave dynamics. Namely, $\overline {T}_{SS} > 1$ at $S=2$ and $K_z
\ll 1$. Physically it means that the amplitude of the reflected
SMW is greater than amplitude of incident SMW. This is the
over-reflection phenomenon first discovered by Miles for acoustic
waves \cite{M57}. This phenomenon becomes dominant at $S \gg 1$.

Figures \ref{fig:fig7} and \ref{fig:fig8} show the case when
$\beta = 100$ and $S = 8,12,16,20$. The dashed lines correspond to
$S = 8$. For this range of the parameters all the wave modes are
coupled. The wave over-reflection phenomenon is more profound.
Particularly,  in Fig. \ref{fig:fig7} ${T}_{SS}, \overline
{T}_{SS} \approx 7$ when $K_z \ll 1$. In other words, the energy
density of the reflected (as well as transmitted) SMW is about 50
times greater that the energy density of the incident SMW. When
$K_z \sim 1$, then ${T}_{AS}, \overline {T}_{AS} \approx 5$, i.e.,
the reflected and transmitted AW are substantially greater then
incident SMW. Figure \ref{fig:fig8} represents the transformation
coefficients if only AW exists initially. In this case significant
growth of energy density of perturbations takes place at $K_z
\gtrsim 1$.

For more detailed analysis of the over-reflection phenomena
consider the dynamics of SMW at $K_z=0,~ \beta \gg 1$ and $S \ll
\beta^{1/2}$. Due to the condition $K_z=0$ there is no coupling
between SMW and AW. Condition  $S \ll \beta^{1/2}$ provides that
the coupling between SMW and FMW is also negligible. Combining
Eqs. (\ref{eq:o8}), and (\ref{eq:aa3})-(\ref{eq:aa5}) one can
obtain following second order equation for the evolution of SMW:
\begin{equation}  \frac{{\rm d^2} \Psi_S}{{\rm d} \tau^2} + f(S,\tau) \Psi_S
 = 0. \label{eq:o18}
\end{equation}
where:
\begin{equation}  f(S,\tau) \equiv 1 - \frac{S^2}{[1+K_y^2(\tau)]^2}
 \label{eq:o181}
\end{equation}

Asymptotic analysis of this kind of equations is well known in
quantum mechanics \cite{LL,K} and the theory of differential
equations \cite{O,F}. From mathematical point of view the
reflection is caused due to the cavity of $f(S,\tau)$ that appears
in the vicinity of the point $\tau_+$ where $K_y(\tau_+)=0$.

In accordance with quantum mechanics we define reflection and
transmission coefficients as $R \equiv {\overline{T}_{SS}^{2}}$
and $T \equiv {T_{SS}^{2}}$ respectively. It is well known
\cite{LL,O,F} that wave action conservation provides that:
\begin{equation}  1 = T_{SS}^2 - \overline{T}_{SS}^{2} \equiv T - R. \label{eq:in3}
\end{equation}
So we can conclude that the total energy of SMW always increase
during the reflection/over-reflection process:
\begin{equation} \frac{E}{E_0} = T + R.\label{eq:e1}
\end{equation}

The dependence of reflection coefficient $R$ on $S$  at $\beta =
400$ obtained by numerical solution of Eqs.
(\ref{eq:o3})-(\ref{eq:o5}) is presented in Fig. \ref{fig:fig9}.
Initial conditions for numerical solution are chosen by WKB
solutions (\ref{eq:wkb}) far on the left side of resonant area
($K_y(0) \gg 1,S$). For high shear rates ($ S \sim \beta^{1/2}$)
SMW is partially transformed to FMW, that is why the growth rate
of the reflection coefficient decreases (dashed part of the graph
in Fig. \ref{fig:fig9}). According to the numerical study of Eq.
(\ref{eq:o18}), the amplitude of reflected SFH exceeds the
amplitude of incident SFH if:
\begin{equation} S^2 \equiv \left( \frac{A}{k_xV_A}\right)^2 > 2.\label{eq:co6}
\end{equation}
This condition indicates that the phenomenon of over-reflection
can take place in plasma with $\beta \gg 1$ even for small values
of the shear parameter $A$. The amplification of the energy
density of perturbations is always finite, but it can become
arbitrarily large under due increase of the shear rate.

\section{Summary and discussions}

The linear dynamics of MHD wave modes in constant shear flows is
studied qualitatively and quantitatively in the framework of
non-modal approach. Usage of asymptotic analysis familiar in
quantum mechanics allows to study the physics of the
(over)reflection and mutual transformation phenomena of MHD wave
modes. Quantitative asymptotic and numerical analysis is performed
for wide range of the system parameters: relation of spanwise and
streamwise wave numbers ($K_z \equiv k_z/k_x$), plasma beta
($\beta \equiv c_s^2/V_A^2$) and the shear rate ($S \equiv A/(k_x
V_A)$). The transformation takes place at small as well as at high
shear rates and involves all MHD wave modes. The over-reflection
becomes apparent only for slow magnetosonic and Alfv\'en wave
modes at high shear rates.

At $S \ll 1$, the transformation has resonant nature and different
wave modes are coupled at different $\beta$.\\ -- In high $\beta$
regime Alfv\'en and slow magnetosonic waves are coupled and the
transformation is effective at $K_z \simeq 1$. At $\beta S \ll 1$,
for the maximum transformation coefficient we have
$(T_{AS}^2)_{max} =1/2$. If $\beta S$ is not small,
$(T_{AS})_{max}=1$, i.e., the total transformation of one wave
mode into another is possible.\\ -- In low $\beta$ regime Alfv\'en
and fast magnetosonic waves are coupled. The transformation is
effective also at $K_z \simeq 1$. $({T_{FA}}^2)_{max} = 1/2$,
i.e., even in the optimal regime only half of the energy of SFH of
FMW can be transformed into AW and vice versa. \\ -- At $\beta
\sim 1$ all of the MHD wave modes are coupled: at $K_z \ll 1$
mainly are coupled slow and fast magnetosonic waves. Whereas at
$K_z \simeq 1$ the coupling of these waves with Alfv\'en wave is
dominant.

At moderate and high shear rates ($S$), the wave mutual
transformation is accompanied by wave (over)reflection phenomenon.
The amplification of the energy density of perturbations is always
finite. However, the ratio of final to initial energies of waves
can become arbitrarily large with increase of the shear rate.

Described phenomena could have substantial applications in the
theory of MHD turbulence that is intensively developing recently
(see, e.g., Ref. \cite{LG03} and references therein). This
concerns both the fundamental problem of MHD turbulence and wave
modes participation in the turbulent processes.

Fundamental problem is the elaboration of the concept of
self-sustaining MHD turbulence. None of the previous studies were
concerned on energy source of the turbulence. The usual procedure
is the following \cite{LG03}: some energy source of perturbations
is assumed {\it a priori} and the evolution of energy spectrum is
studied. On the other hand, in the framework of canonical MHD
theory, there is no linear instability in the flows (they are
spectrally stable). Therefore, existence of an energetic source
for the permanent forcing of turbulence is problematic in the
canonical/spectral theory. Nevertheless, one can apply the concept
of bypass turbulence elaborated by hydrodynamic community for
spectrally stable shear flows. The role of the energy source in
the MHD flows could play waves non-exponential growth
(over-reflection).

Due to the importance of over-reflection phenomenon as a possible
energy source for MHD turbulence in spectrally stable flows,
detailed analysis of evolution of SFHs of MHD wave modes in flows
with high shear rate will be treated more extensively elsewhere.

As for ingredients of the turbulence: the study of homogeneous MHD
turbulence provides that in the case of $\beta \gg 1$, the
turbulence is mainly Alfv\'enic (i.e., mainly consists of AW) and
SMW plays a role of passive admixture. Studied in this paper
linear coupling of MHD wave modes can significantly change this
scenario even in the presence of small velocity shear ($S \ll 1$).
Thus for the wide range of the system parameters all of the wave
modes (at list two of them) should be the ingredients of the real
MHD turbulence -- the turbulence should be of a "mixed" type.

\begin{acknowledgments}

This research is supported by GRDF grant 3315 and GAS grant
2.39.02.

The authors are immensely grateful to Alexander Tevzadze and Temur
Zaqarashvili for valuable help in the preparing of the final
version of this article.

\end{acknowledgments}

\appendix

\section{coupling coefficients}

Substituting Eq. (\ref{eq:mx1}) into Eqs.
(\ref{eq:aa1})-(\ref{eq:aa2}) after straightforward calculations
one can obtain:
\begin{equation}
{\bf \Theta} = \left( \begin{array}{ccc} {\dot{\alpha}}^2
+{\dot{\gamma}}^2 \sin^2{\alpha}  \\ {\dot{\alpha}}^2
+{\dot{\gamma}}^2 \cos^2{\alpha} \\ {\dot{\gamma}}^2
\end{array} \right),\label{eq:aa3}
\end{equation}
\begin{equation}
{\bf \Lambda} = \left( \begin{array}{ccc} 0 & \dot{\alpha} &
\dot{\gamma} \sin{\alpha}  \\ -\dot{\alpha} & 0 & -\dot{\gamma}
\cos{\alpha}
\\ -\dot{\gamma} \sin{\alpha} & \dot{\gamma} \cos{\alpha} & 0
\end{array} \right),\label{eq:aa4}
\end{equation}
and for non-zero components of ${\bf \Upsilon}$:
\begin{eqnarray}
\begin{array}{ll}
\Upsilon_{12} = & {\ddot{\alpha}}^2 + \dot{\gamma} \sin{\alpha}
\cos{\alpha},
\\ \Upsilon_{13} = & {\dot{\gamma}}^2 \sin{\alpha}, \\ \Upsilon_{21} = &
-\ddot{\alpha} + \dot{\gamma}^2 \sin{\alpha} \cos{\alpha},
\\ \Upsilon_{23} = &-\ddot{\gamma} \cos{\alpha}, \\
\Upsilon_{31} = & -2\dot{\alpha} \dot{\gamma} \cos{\alpha}
-\ddot{\gamma} \sin{\alpha},
\\ \Upsilon_{32} = & -2\dot{\alpha} \dot{\gamma} \sin{\alpha}  +
\ddot{\gamma} \cos{\alpha},
\end{array}\label{eq:aa5}
\end{eqnarray}
where $\alpha(\tau)$ and $\gamma(\tau)$ are defined by Eqs.
(\ref{eq:alp0}) and (\ref{eq:gam0}) with $K_y(\tau)\equiv
K_y-S\tau$.

\section{Some mathematical aspects of Linear Transformations}

To avoid the complication of mathematical formalism first consider
the case of two coupled oscillators:
\begin{equation}
 \frac{{\rm d^2} \psi_1}{{\rm d} \tau^2} + G_{11} \psi_1=
 G_{12} \psi_2,
 \label{eq:ab0}
\end{equation}
\begin{equation}
 \frac{{\rm d^2} \psi_2}{{\rm d} \tau^2} + G_{22} \psi_2 =
 G_{21}\psi_1,
\label{eq:ab01}
\end{equation}
with $G_{12}=G_{21}$ and eigen-frequencies:
\begin{equation}
\Omega^2_{1,2} = \frac{1}{2} \left[ G_{11} +G_{22} \pm
\sqrt{(G_{11}-G_{22})^2 + 4G_{12}^2} \right], \label{eq:abf}
\end{equation}
that are slowly varying functions of $\tau$:
\begin{equation}
{\dot \Omega}_i \ll \Omega_i^2,~~~~~~~~i=1,2. \label{eq:bwkbc}
\end{equation}
$\tau$ is normalized so that $\Omega_i(S\tau) \sim 1$ and $S \ll
1$. In addition, let us assume that all eigen-frequencies are real
during the evolution. It can be easily seen from Eqs.
(\ref{eq:ab0})-(\ref{eq:ab01}) that this condition implies:
\begin{equation}
G_{11} G_{22} > G_{12}^2. \label{eq:abc}
\end{equation}
From this condition it follows that neither turning points (where
$\Omega_i=0$) nor resonant points ($\Omega_1^2 = \Omega_2^2 $) can
be located on the real $\tau $ axis. Transformation matrix to the
normal variables, similar to Eq. (\ref{eq:mx1}), is:
\begin{equation}
{\bf Q} = \left( \begin{array}{cc} \cos \lambda & \sin \lambda  \\
\sin \lambda & -\cos \lambda
\end{array} \right),
\label{eq:abm}
\end{equation}
with:
\begin{equation}
\lambda \equiv
\arctan\left(\frac{G_{12}}{\Omega_2^2-G_{11}}\right).
\label{eq:ab02}
\end{equation}
In the normal variables the set of equations
(\ref{eq:ab0})-(\ref{eq:ab01}) reduces to:
\begin{equation}
 \ddot{\Psi}_1 + \left( \Omega_1^2 - \dot{\lambda}^2 \right) \Psi_1 =
 - 2 \dot{\lambda} \dot{\Psi}_2 - \ddot{\lambda} \Psi_2, \label{eq:ab1}
\end{equation}
\begin{equation}
 \ddot{\Psi}_2 + \left( \Omega_2^2 - \dot{\lambda}^2 \right) \Psi_2 =
 2 \dot{\lambda} \dot{\Psi}_1 + \ddot{\lambda} \Psi_1,\label{eq:ab2}
\end{equation}
where:
\begin{equation}
\dot{\lambda} \equiv \frac{(\Omega_2^2-G_{11}) dG_{12}/d\tau
-G_{12}d({\Omega^2_2}-{G_{11}})/d\tau
}{(\Omega_2^2-G_{11})(\Omega_2^2-\Omega_1^2)}. \label{eq:ab03}
\end{equation}

According to Eq. (\ref{eq:abf}) condition $\Omega_2^2=G_{11}$
leads to $ G_{12}=0$. Consequently, as it was mentioned above,
only singular points of coupling coefficient are resonant points,
where $\Omega_1^2=\Omega_2^2$.

Let $\bar{\Psi}_{1}^\pm$ and $\bar{\Psi}_{2}^\pm$ be linear
independent solutions of corresponding homogeneous equations (Eqs.
(\ref{eq:ab1})-(\ref{eq:ab2}) with right hand sides equal to
zero), normalized such that asymptotically they converge to WKB
solutions:
\begin{equation}
\bar{\Psi}_{1,2}^\pm \approx \frac{1}{\sqrt{\Omega_{1,2}}}\exp
\left( \pm i \int \Omega_{1,2}d\tau \right).\label{eq:ab3}
\end{equation}
The general solution of the set of Eqs.
(\ref{eq:ab1})-(\ref{eq:ab2}) is ($i=1,2$):
\begin{equation}
\Psi_{i}= D_{i}^+ \bar{\Psi}_{i}^+ + D_{i}^-
\bar{\Psi}_{i}^-.\label{eq:ab311}
\end{equation}

For simplicity let us assume, that far on the left side of the
resonant area only one, for example $ \Psi_{2}^+$ mode exists,
i.e., $D_2^+ =1$ and all other WKB amplitudes are zero.
Considering terms on the right hand side of Eqs.
(\ref{eq:ab1})-(\ref{eq:ab2}) as external source, using well known
expressions for the solution of linear inhomogeneous second order
differential equation \cite{AS} and integrating by parts one can
estimate the amplitude of generated wave $\Psi_1^+$ as:
\begin{equation}
D_1^+ \sim \int_{-\infty}^{\infty} \dot{\lambda} \left(
\bar{\Psi}_{1}^- \frac{d \bar{\Psi}_{2}^+}{d \tau} -
\bar{\Psi}_{2}^+ \frac{d \bar{\Psi}_{1}^-}{d \tau} \right) d
\tau.\label{eq:ab4}
\end{equation}

Necessary condition for effective coupling implies:
\begin{equation}
D_1^+ \sim 1.\label{eq:ab5}
\end{equation}

Presenting $ \Psi_{1}^-$ and $ \Psi_{2}^+$ in the form of formal
series \cite{O}, substituting them into Eq. (\ref{eq:ab4}) and
taking into account that obtained series of integrals converges
rapidly \cite{A60,B90}, condition (\ref{eq:ab5}) takes the form:
\begin{equation}
\int \dot{\lambda} \exp \left( {i \int^\tau (\Omega_1-\Omega_2)d
\tau_1} \right) d\tau \sim 1.\label{eq:ab6}
\end{equation}
If $S \ll 1 $, then phase multiplier in (\ref{eq:ab6}) is rapidly
oscillating. Condition (\ref{eq:ab5}) can be satisfied only if
there exists an interval where time scale of the changing $
\dot{\lambda}$ and the phase multiplier in the integrand of Eq.
(\ref{eq:ab6}) are at least the same. This lead to the condition:
\begin{equation}
|\Omega_1^2-\Omega_2^2| \leq |\dot{\lambda}
\Omega_2|,\label{eq:ab7}
\end{equation}
at some interval along the path of integration. On the other hand
one can note that only singular points of $ \dot{\lambda}$ in the
complex $ \tau$-plane are resonant points (where $
\Omega_1=\Omega_2$). If so, Landau's rule \cite{LL} provides that
if $ S \ll 1$, then leading term of asymptotics of transformation
coefficient is:
\begin{equation}
T_{12} \sim \exp \left( - \left| {\rm Im}
\int_{\tau_0}^{\tau_{12}} (\Omega_1-\Omega_2)d t \right| \right),
\label{eq:ab8}
\end{equation}
where $ \tau_0$ is arbitrary point on the real $\tau$-axis and
$\tau_{12}$ is the nearest to the real $\tau$-axis resonant point.
Eq. (\ref{eq:ab8}) represents an alternative necessary condition
for effective coupling of wave modes. Namely, resonant point
$\tau_{12}$ must be close to the real $ \tau$-axis in such a way
that the exponent in right hand side of Eq. (\ref{eq:ab8}) has to
be of the order of unity. Surely this condition is equivalent to
Eq. (\ref{eq:ab7}). On the other hand, it shows that if the
condition of effective coupling is not satisfied, then the
transformation coefficient is exponentially small with respect to
the large parameter $1/S$.

According to Eq. (\ref{eq:ab03}), in the neighborhood of $n$th
order resonant point:
\begin{equation}
\dot{\lambda} \sim \frac{S}{(S\Delta
\tau_{12})^{n/2}}.\label{eq:ab9}
\end{equation}
Eq. (\ref{eq:ab7}) yields following estimation for the effective
coupling time scale:
\begin{equation}
\Delta \tau_{12} \sim S^{-1+1/n}. \label{eq:ab91}
\end{equation}

\subsection{Resonant points of the first order}

Consider the case when only a pair of complex conjugated first
order resonant points exists close to the real $ \tau$-axis. This
means that in the neighborhood of $ \tau_{12}$:
\begin{equation}
|\Omega_1^2 - \Omega_2^2| \sim (\tau-\tau_{12})^{1/2}.
\end{equation}
In this case, the leading term of asymptotic of transformation
coefficient can be found exactly by the method first used by
Landau \cite{LL}.

Consider Eqs. (\ref{eq:ab1})-(\ref{eq:ab2}) along the path $\zeta$
that turns over $ \tau_{12}$ in the complex $\tau$-plane
(Fig.\ref{fig:fig10}). Considering Eqs.
(\ref{eq:ab1})-(\ref{eq:ab2}) in the neighborhood of $\tau_{12}$,
where $\dot{\lambda} \sim (\tau-\tau_{12})^{-1/2}$, one can obtain
that Eqs. (\ref{eq:ab1}) and (\ref{eq:ab2}) exchange after turning
over resonant point. This means that $\Psi_2$ became $ \Psi_1$ and
vise versa. After this kind of analysis it is usually concluded
(see e.g., Ref. \cite{LL}), that the transformation coefficient
is:
\begin{equation}
T_{12} = \exp \left( - \left | {\rm Im} \int_{\tau_0}^{\tau_{12}}
( \Omega_1 - \Omega_2) d \tau \right | \right)
[1+O(S^{1/2})].\label{eq:ab92}
\end{equation}

From our point of view this kind of analysis needs to be
commented. When one considers the set of Eqs.
(\ref{eq:ab1})-(\ref{eq:ab2}) along the path $\zeta$, except
mentioned above exchange of $\Psi_{1}$ and $\Psi_{2}$, there are
also terms
\begin{equation} \int_\zeta
\dot{\lambda} \left( \bar{\Psi}_{1}^-\frac{d \bar{\Psi}_{2}^+}{d
\tau} - \bar{\Psi}_{2}^+ \frac{d \bar{\Psi}_{1}^-}{d \tau} \right)
d \tau. \label{eq:abl}
\end{equation}
In the case of the first order resonant points it can be shown
that according Watson's lemma contribution of these terms are of
order $S^{1/2}
\exp(-|\int_{\tau_0}^{\tau_{12}}(\Omega_1-\Omega_2)d\tau|)$ and
(\ref{eq:ab92}) is valid. However, in the case of the second or
higher order resonant points, the contribution of terms like
(\ref{eq:abl}) is at least of the same order as (\ref{eq:ab92}).
That is why this asymptotic method works only in the case of first
order resonant points.

\subsection{Resonant points of the second order}

Unfortunately the transformation coefficient can not be obtained
analytically in the case of the second or higher order resonant
points. Exact asymptotic expressions can be obtained only in the
case of weak interaction. Let us consider the case when $
\tau_{12}$ and $ \tau_{12}^\ast$ are complex conjugated second
order resonant points. Assume again, that far away on the left
side of the resonant area only one, say, $\Psi_{2}^+$ mode exists.
In the case of a weak interaction feedback of $\Psi_{1}^+$ to
$\Psi_{2}^+$ in Eqs. (\ref{eq:ab1})-(\ref{eq:ab2}) can be
neglected, so these equations decouple  and there remains only
inhomogeneous second order differential equation. We skip some
details of calculations noting only main steps. In the case of
neglible feedback, for the transformation coefficient one can
obtain:
\begin{equation}
T_{12} = \frac{1}{2} \int_{-\infty}^{\infty} \dot{\lambda} \left(
\bar{\Psi}_{1}^-\frac{d \bar{\Psi}_{2}^+}{d \tau} -
\bar{\Psi}_{2}^+ \frac{d \bar{\Psi}_{1}^-}{d \tau} \right) d \tau.
\end{equation}
In the neighborhood of the resonant point one can use the formal
asymptotic series for $\Psi_{1,2}^\pm $, considering
$\Omega_{1,2}(\tau) = \Omega_{1,2}(\tau_{12}) $ as constants.
Consequently, the last equation takes the form:
\begin{equation}
T_{12} = \left | \Sigma_n \int_{-\infty}^{\infty} \chi_n
{\dot{\lambda}}^n \exp\left(i \int^\tau (\Omega_1-\Omega_2) d t
\right) d \tau \right |,\label{eq:ab10}
\end{equation}
where $ \chi_n$ are regular functions in the neighborhood of $
\tau_{12}$ and:
\begin{equation}
\chi_1 = \frac{\Omega_1+\Omega_2}{2\sqrt{\Omega_1 \Omega_2}}.
\end{equation}

To evaluate the integrals in (\ref{eq:ab10}), one can use Van der
Wearden's method \cite{L}. It is not difficult to show that each
term in Eq. (\ref{eq:ab10}) are of order $S^{1/2}$ with respect to
the previous term. Consequently, in the case of the second order
resonant point, leading term of the asymptotics is defined by the
first term, for which we have:
\begin{equation}
T_{12} = \frac{\pi}{2}  \exp \left( - \left | {\rm Im}
\int_{\tau_0}^{\tau_{12}} (\Omega_1 - \Omega_2)  d \tau \right |
\right) .\label{eq:ab11}
\end{equation}

\newpage

Figure captions

\begin{figure}[t]
\includegraphics[]{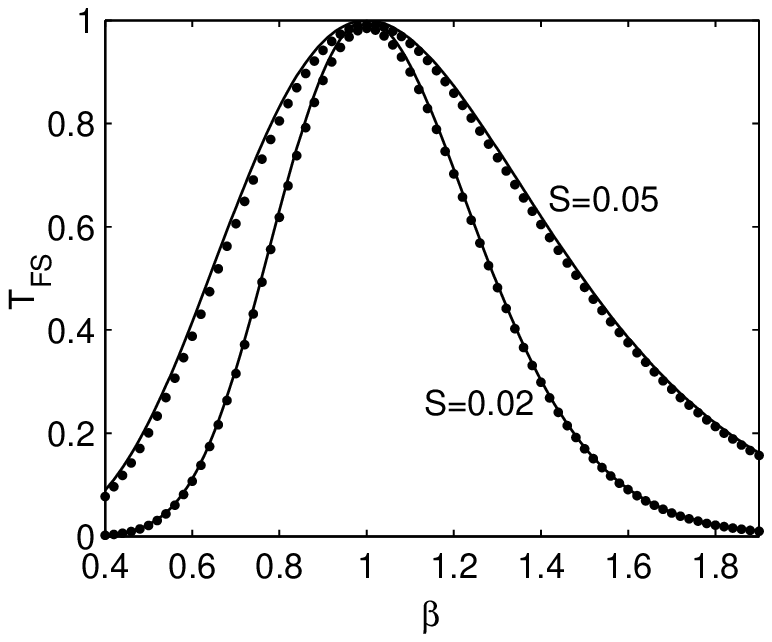}
\caption{\label{fig:fig1}}
\end{figure}

FIG. \ref{fig:fig1}: Transformation coefficient $T_{FS}$ vs
$\beta$ for $ S=0.05$ and $S=0.02$ obtained by numerical (dotted
line) and analytical solutions (solid line), see Eq.
(\ref{eq:tr6}).

\begin{figure}[t]
\includegraphics[]{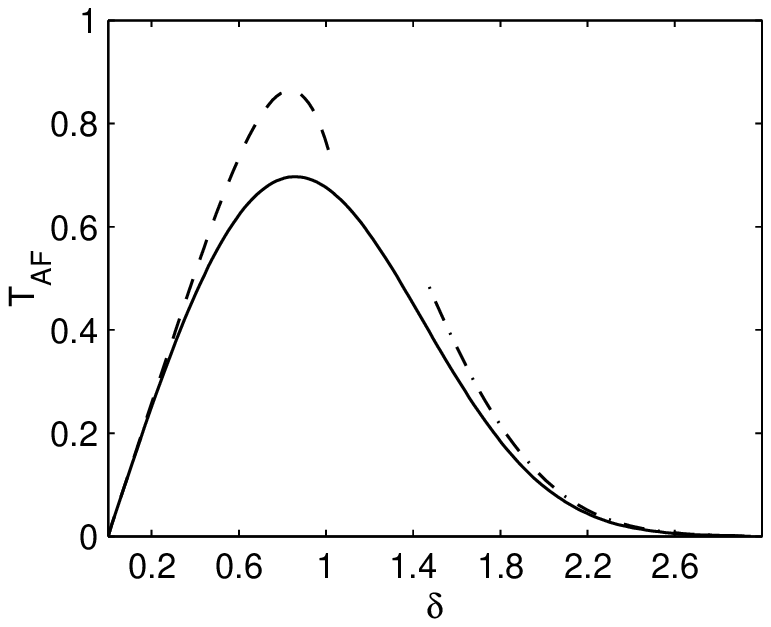}
\caption{\label{fig:fig2}}
\end{figure}

FIG. \ref{fig:fig2} Transformation coefficient $ T_{AF}$ vs
$\delta$. Dash-dotted line and dashed line represent analytical
expressions (\ref{eq:tr8}) and (\ref{eq:tr10}), respectively.
Solid line is obtained by numerical solution of Eqs.
(\ref{eq:o14})-(\ref{eq:o15}).

\begin{figure}[t]
\includegraphics[]{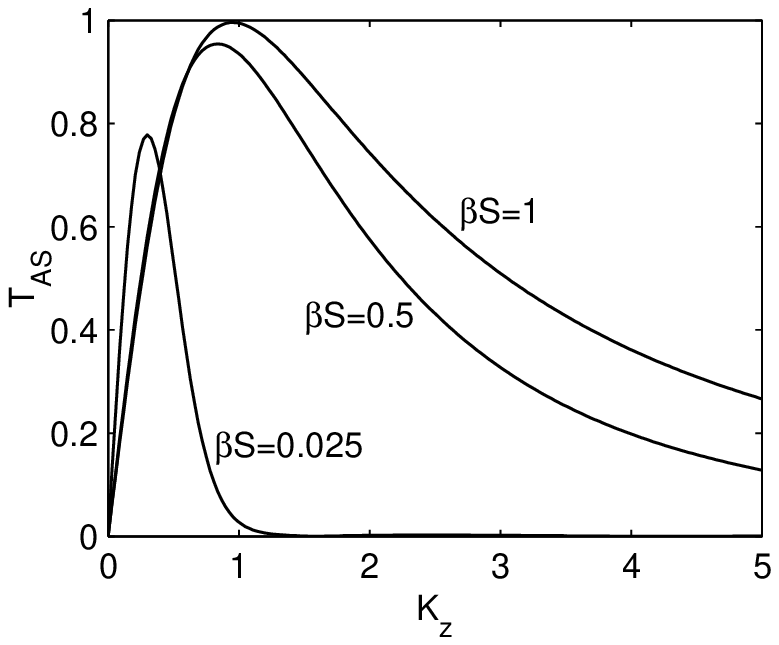}
\caption{\label{fig:fig3}}
\end{figure}

FIG. \ref{fig:fig3} Transformation coefficient $T_{AS} $ vs $K_z $
for $S=0.005$ and $\beta=5~(\beta S = 0.025)$, $\beta=100~(\beta S
= 0.5)$ and $\beta=200~(\beta S = 1)$.

\begin{figure}[t]
\includegraphics[]{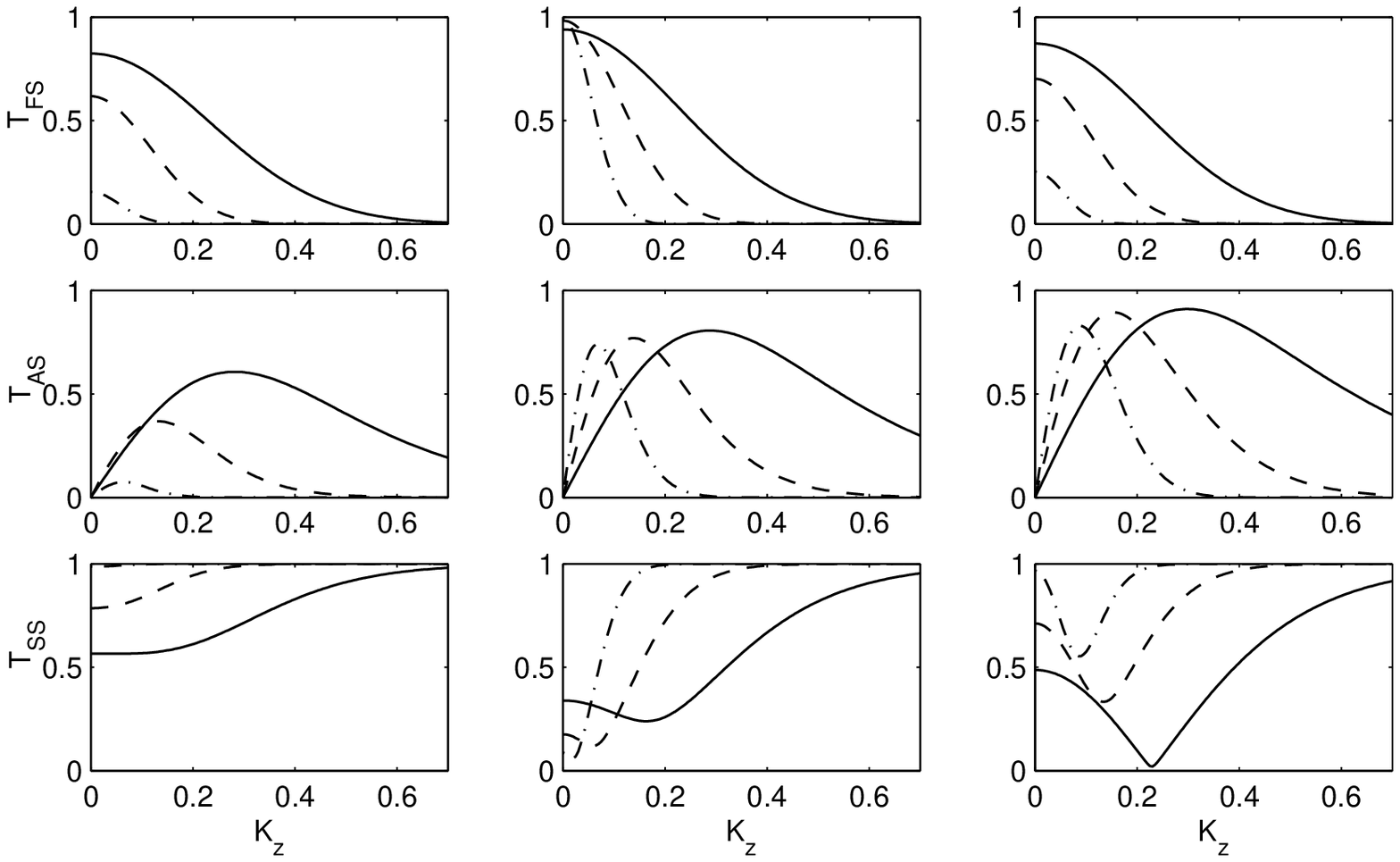}
\caption{\label{fig:fig4}}
\end{figure}

FIG. \ref{fig:fig4} Transformation coefficients $T_{FS}$, $T_{AS}$
and $T_{SS}$ vs $K_z $ for $\beta=0.8$ (left column), $\beta=1.0$
(center column) and $\beta=1.2$ (right column) for different
normalized shear parameters: $S=0.005$ (dash-dotted lines),
$S=0.02$ (dashed lines) and $S=0.08$ (solid lines). Note, that
according to the conservation of wave action
$T_{FS}^2+T_{AS}^2+T_{SS}^2=1$ (see Eq. \ref{eq:in1}).

\begin{figure}[t]
\includegraphics[]{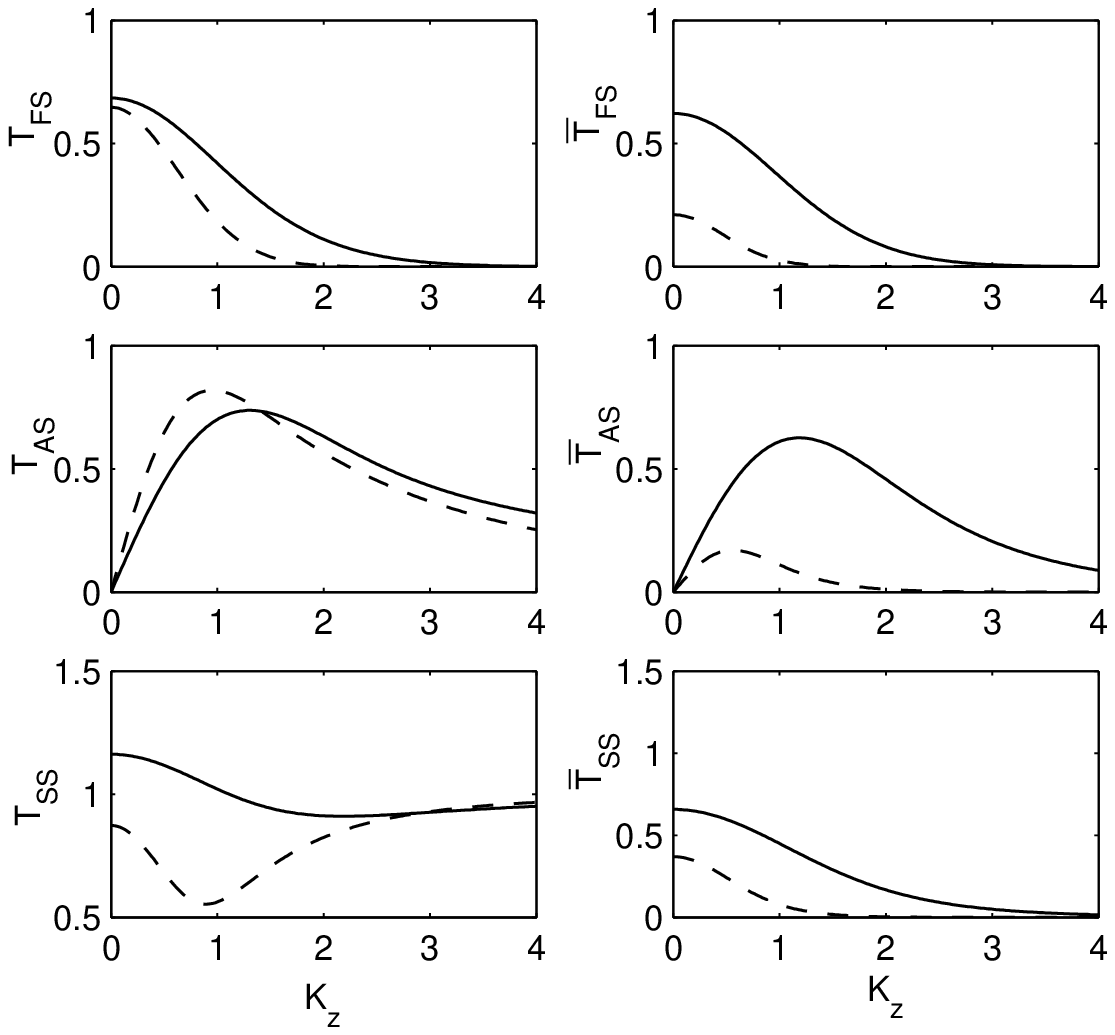}
\caption{\label{fig:fig5}}
\end{figure}

FIG. \ref{fig:fig5} Transformation coefficients ($T_{iS} $ and
$\overline T_{iS} $ respectively) vs $K_z$ at $\beta=2$ at
different shear rates: $S = 1$ (dashed lines) and $S = 4$ (solid
lines).

\begin{figure}[t]
\includegraphics[]{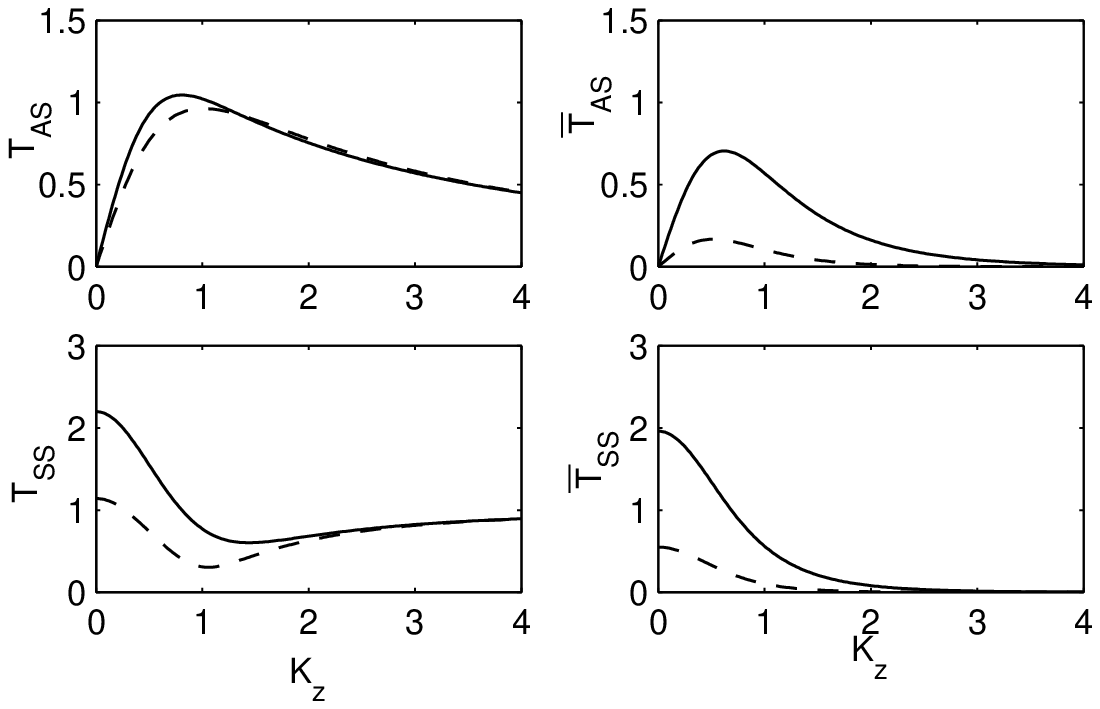}
\caption{\label{fig:fig6}}
\end{figure}

FIG. \ref{fig:fig6} Transformation coefficients
$T_{AS}$,$\overline T_{AS}$, $T_{SS}$ and $\overline T_{SS}$ vs
$K_z $ for $\beta=50$ at different shear rates: $S=1.0$ (dotted
lines) and $S=2.0$ (solid lines). $T_{FS}, \overline T_{FS}
\approx 0$.

\begin{figure}[t]
\includegraphics[]{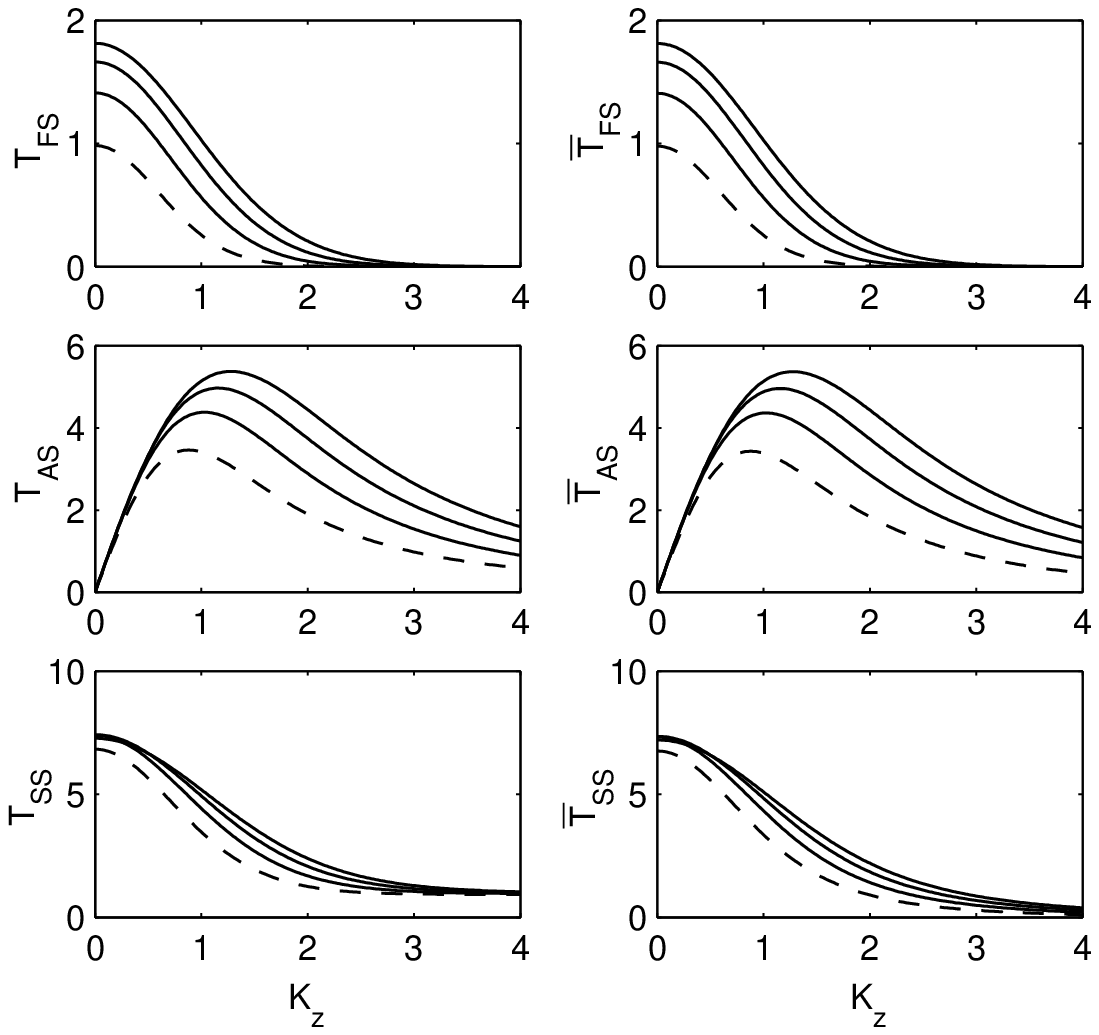}
\caption{\label{fig:fig7}}
\end{figure}

FIG. \ref{fig:fig7} Transformation coefficients ($T_{iS} $ and
$\overline T_{iS} $ respectively) vs $K_z$ at $\beta=100$ at
different shear rates: $S = 8,12,16,20$. Dashed lines correspond
to $S=8$.

\begin{figure}[t]
\includegraphics[]{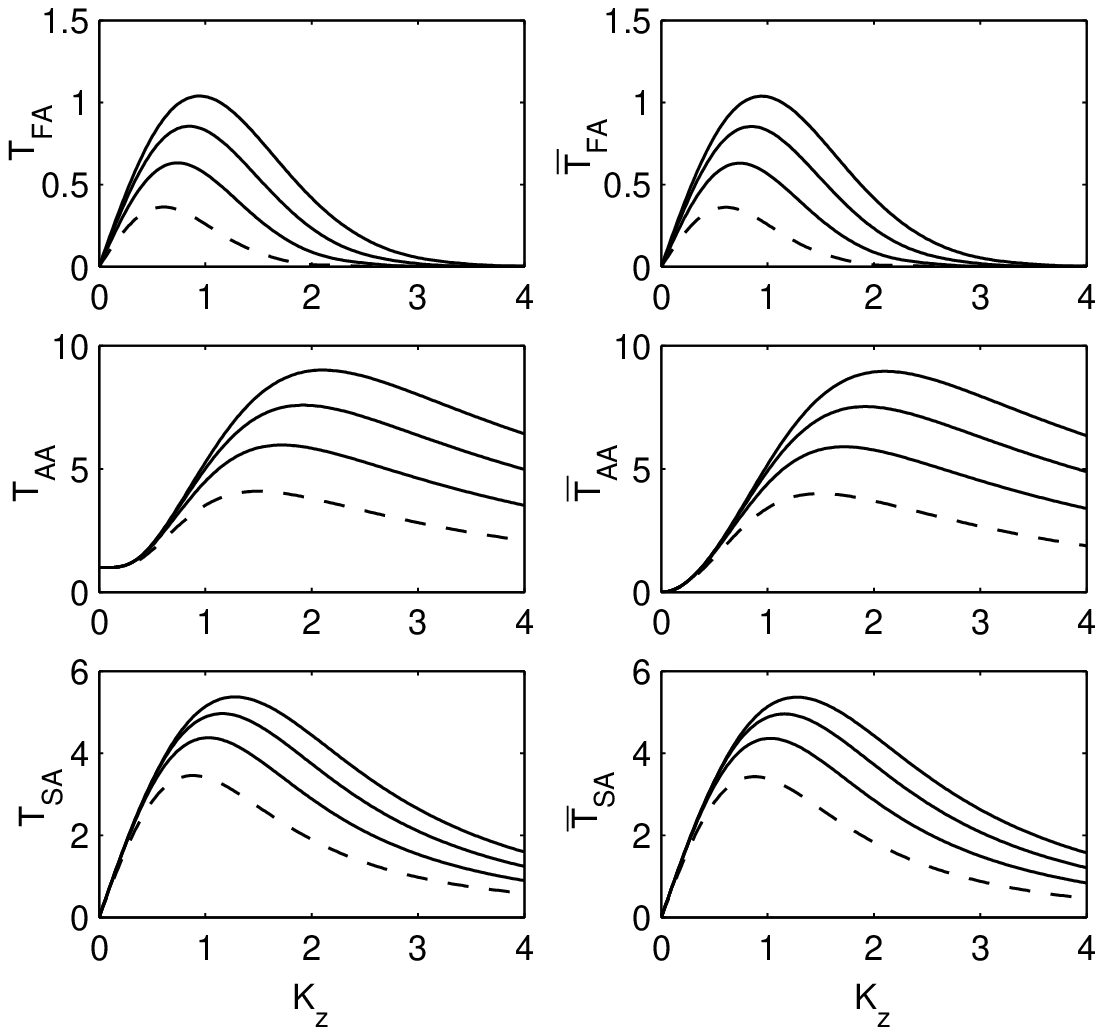}
\caption{\label{fig:fig8}}
\end{figure}

FIG. \ref{fig:fig8} Transformation coefficients ($T_{iA} $ and
$\overline T_{iA}$ respectively) vs $K_z$ at $\beta=100$ and
different shear rates: $S = 8,12,16,20$. Dashed lines correspond
to $S=8$.

\begin{figure}[t]
\includegraphics[]{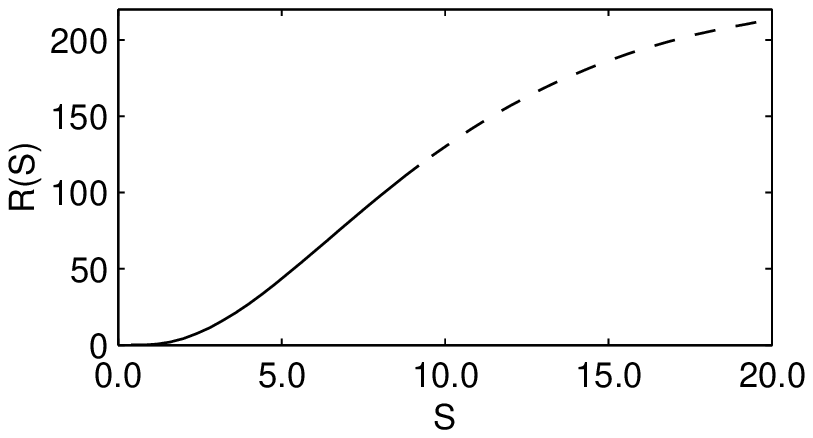}
\caption{\label{fig:fig9}}
\end{figure}

FIG. \ref{fig:fig9} Reflection coefficient of SMW $R(S)$ vs
normalized shear parameter $S$. Here $ \beta=400$. For high shear
parameters ($ S \sim \beta^{1/2} $) SMW is partially conversed to
FMW, that is why that slope of the refraction coefficient graph
decreases (dashed part of the graph).

\begin{figure}[t]
\includegraphics[]{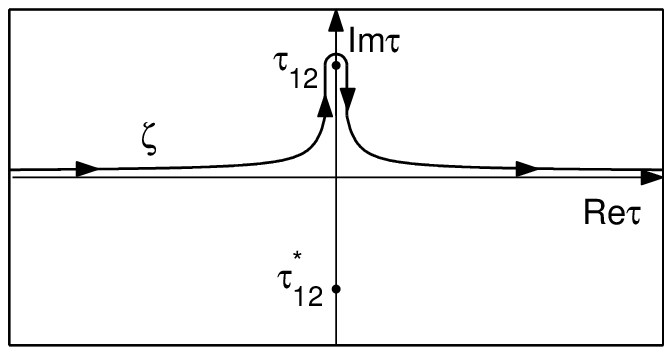}
\caption{\label{fig:fig10}}
\end{figure}

FIG. \ref{fig:fig10} Resonant points and path $ \zeta $ in the
complex $ \tau $ plane.

\newpage

\end{document}